\documentclass[fleqn,usenatbib,useAMS]{mn2e}  
\usepackage{times}
\usepackage{epsf}
\usepackage{amssymb}

\title[The metal enrichment of the ICM]{
The metal enrichment of the 
intracluster medium in hierarchical galaxy formation models 
}

\author[M. Nagashima et al.]{Masahiro Nagashima
\thanks{E-mail: masa@scphys.kyoto-u.ac.jp (MN).
}
\thanks{Present address: Department of Physics, Kyoto University,
Sakyo-ku, Kyoto 606-8502, Japan},
Cedric G. Lacey,
Carlton M. Baugh,
Carlos S. Frenk,
\newauthor{Shaun Cole}
\\ 
Department of Physics, University of Durham, South Road, Durham DH1 3LE,
United Kingdom
}

\begin{document}   
\maketitle   

\begin{abstract}   
We investigate the metal enrichment of the intracluster medium (ICM)
in the framework of hierarchical models of galaxy formation. We
calculate the formation and evolution of galaxies and clusters using a
semi-analytical model which includes the effects of flows of gas and
metals both into and out of galaxies. For the first time in a
semi-analytical model, we calculate the production of both $\alpha$
and iron-peak elements based on theoretical models for the lifetimes
and ejecta of type Ia and type II supernovae (SNe~Ia and SNe~II). It
is essential to include the long lifetimes of the SNIa progenitors in
order to correctly model the evolution of the iron-peak elements. We
find that if all stars form with an IMF similar to that found in the
solar neighbourhood, then the metallicities of O, Mg, Si and Fe in the
ICM are predicted to be 2-3 times lower than observed values. In
contrast, a model (also favoured on other grounds) in which stars
formed in bursts triggered by galaxy mergers have a top-heavy IMF
reproduces the observed ICM abundances of O, Mg, Si and Fe. The same
model predicts ratios of ICM mass to total stellar luminosity in
clusters which agree well with observations. According to our model,
the bulk of the metals in clusters are produced by $L_{*}$ and
brighter galaxies. We predict only mild evolution of [Fe/H] in the ICM
with redshift out to $z\sim 1$, consistent with the sparse data
available on high-z clusters.  By contrast, the [O/Fe] ratio is
predicted to gradually decrease with time because of the delayed
production of iron compared with oxygen.  We find that, at a given
redshift, the scatter in global metallicity for clusters of a given
mass is quite small, even though the formation histories of individual
clusters show wide variations. The observed diversity in ICM
metallicities may thus result from the range in metallicity gradients
induced by the scatter in the assembly histories of clusters of
galaxies.

\end{abstract}   
   

\section{INTRODUCTION}   

The bulk of the baryons in galaxy clusters are in their hot gas
component, the intracluster medium (ICM), which typically contains
5-10 times the mass found in cluster galaxies.  The metal content of
the ICM is sensitive to the star formation histories of the cluster
galaxies and to the way in which metals are ejected from galaxies in
supernova explosions and winds.  In particular, the abundances of
different species can be used to constrain the relative numbers of
type Ia (SNe~Ia) and type II (SNe~II) supernovae that occurred in the
cluster galaxies.  This is because the abundances of the $\alpha$
elements (e.g. O, Mg) are driven primarily by the rate of SNe~II,
whereas the production of iron is dominated by SNe~Ia.  The abundance
of $\alpha$ elements in the ICM is found, from X-ray measurements, to
be comparable to abundances in the solar neighbourhood, whereas the
abundance of iron is only about 30\% of that found locally
\citep[e.g.][]{m96}.  This discrepancy has led several authors to
propose that the initial mass function (IMF) of stars forming in
cluster galaxies could be substantially different from that inferred
for star formation in less dense environments
\citep[e.g.][]{renzini93, zs96, m96, v03, tbmrt04, mg04}.

The chemical evolution of galaxies has been modelled extensively in the
monolithic collapse scenario \citep[e.g.,][]{l69}.  Simple infall models
with delayed metal enrichment due to SNe~Ia have been used to
demonstrate that the metallicity distribution of stars in the solar
neighbourhood can be explained by bottom-heavy IMFs as proposed by
\citet{s55} \citep{mg86, mf89, tnyhyt95, pt95, ytn96}.  The metal
enrichment of the ICM has also been studied in the framework of
monolithic collapse, with the conclusion reached that Salpeter-like IMFs
produce too few heavy elements to match the observed metal abundances in
the ICM \citep{dfj91, mg95, lm96, gm97a, gm97b, mpc03}.  To reconcile
the predictions with the observed ICM abundances, these studies
concluded that a flatter IMF than Salpeter is required, such as the
Arimoto-Yoshii IMF, with $x\simeq 1$ \citep{ay87}, where the number of
stars per logarithmic interval in mass is given as ${\rm d}N/{\rm d}\ln
m\propto m^{-x}$. (For comparison, the Salpeter IMF has $x=1.35$.)  This
conclusion has been confirmed by numerical simulations \citep{v03, tbmrt04}.

There is now overwhelming evidence supporting the gravitational
instability paradigm for the formation of structure in the Universe
\citep[e.g.][]{peacock01, baugh04a}.  Any realistic model of galaxy
formation should therefore be placed in the proper context of a model
that follows the hierarchical formation of cosmic structures, such as
the successful $\Lambda$-dominated cold dark matter ($\Lambda$CDM)
model \citep{percival02, wmap03}.  Semi-analytic models of galaxy
formation provide a framework within which the physics of galaxy
formation can be followed at the same time as dark matter halos are
built up through mergers and the accretion of smaller objects
\citep[e.g.,][]{wf91, c91, kwg93, cafnz94, clbf00, sp99, ntgy01,
nytg02, mcfgp02, hatton03}.  These models are able to reproduce a wide
range of properties of the local galaxy population and to make
predictions for the high redshift universe
\citep[e.g.][]{baugh04b}. 
To date, the majority of semi-analytic models have used the
instantaneous recycling approximation for chemical enrichment, and
only considered metal production by SNe~II. Such models have been used
to investigate the stellar and gas metallicities of disk and
elliptical galaxies and of damped Ly$\alpha$ systems, with broad
agreement obtained between model predictions and observations
\citep{k96, kc98, clbf00, spf01, ny03, ongy04}. Semi-analytic models
based on instantaneous recycling have also been applied to enrichment
of the ICM by Kauffmann \& Charlot (1998) and \cite{del04}, as well as
by Nagashima \& Gouda (2001). The former two papers find that the
observed ICM metallicities are only reproduced in models in which the
chemical yield is twice the standard value for a solar neighbourhood
IMF, in agreement with the conclusions from monolithic collapse models.

The first attempt to take into account metal enrichment due to SNe~Ia
in hierarchical galaxy formation models was that of \citet{t99} and
\citet{tk99}. However, their calculations used only the star
formation histories for spiral and elliptical galaxies extracted from
the Munich group's semi-analytical model, and assumed closed-box
chemical evolution. They did not include the effects of gas inflows
and outflows on the chemical enrichment histories of galaxies.  The
metal enrichment due to SNe~Ia was consistently included in a
semi-analytical model for the first time by \citet{no03}, who took
into account the recycling of metals between stars, the cold gas in
the disk and the hot halo gas.  While the lifetime of SNe~Ia's was
simply assumed to be a constant 1.5Gyr, their model provides good
agreement with the observed distributions of [Fe/H] and of [O/Fe]
vs. [Fe/H] for both solar neighbourhood and bulge stars, based on a
Salpeter-like IMF.

Recently, \citet{baugh04b} showed that the observed number counts of
faint sub-mm sources can be matched in hierarchical models if
starbursts triggered by galaxy mergers are assumed to take place with
a top-heavy IMF (with slope $x=0$).  Since varying the slope of the
IMF affects the abundance ratios of $\alpha$ elements to iron, an
important test of this model is to examine its predictions for the
metallicity of the ICM.  In this paper, using essentially the same
model as used by \citet{baugh04b}, we investigate the metal abundances
of the ICM.  We adopt a standard, ``bottom-heavy'' IMF proposed by
\citet{k83}, with $x=1.5$ for $m\geq 1M_{\odot}$, for quiescent star
formation in disks, and a top-heavy IMF with $x=0$ in starbursts.
While the IMF for quiescent star formation is reasonably well
determined from observations in the solar neighbourhood
\citep[e.g.][]{k02}, the form of the IMF appropriate to starbursts is
still unclear.  There is, however, some tentative observational
evidence supporting a top-heavy IMF \citep{takp97, sg01}.

In this paper, we incorporate metal enrichment due to SNe~Ia as well
as SNe~II into the {\sc galform} semi-analytic galaxy formation model
\citep{clbf00}, to see whether the model with a top-heavy IMF is able
to reproduce the metal abundances of the ICM.  The model includes both
inflows of gas and metals into galaxies due to gas cooling in halos,
and ejection of gas and metals by galactic winds. We calculate the
production of different elements by SNe~Ia and SNe~II based on
theoretical yields as a function of stellar mass, integrated over the
model IMF. The lifetimes of SNe~Ia are consistently calculated for the
adopted stellar IMFs, while instantaneous recycling is assumed for
SNe~II. We compute ICM abundances for O, Mg, Si and Fe for clusters of
different masses, and compare these with observational data. In making
these comparisons, we correct the observed abundances for radial
metallicity gradients; these corrections are important for Si and Fe.

The paper is set out as follows.  In Section~2, we briefly describe the
{\sc galform} model.  In Section~3, we provide a detailed explanation of
how we model metal enrichment. In Section~4, we show what our model
predicts for the present-day luminosity function of galaxies, for
different assumptions about feedback and thermal conduction.  In
Section~5, we present our predictions for the abundances of different
elements in the ICM, and for the ratios of ICM mass to stellar
luminosity and to total mass of dark halos, as functions of cluster
X-ray temperature, and compare with observational data on present-day
clusters. Section~6 compares the model with observational data on the
evolution of ICM abundances with redshift. Section~7 examines the
predicted metal enrichment histories of clusters in more
detail. Finally, Section~8 presents our conclusions.  The Appendix
describes how we correct the observational data for radial abundance
gradients, to estimate global ICM metallicities.

Throughout this paper, the cosmological parameters are fixed to be
$\Omega_{0}=0.3, \Omega_{\Lambda}=0.7, h\equiv H_{0}/100$
km~s$^{-1}$Mpc$^{-1}$ = 0.7 and $\sigma_{8}=0.93$.  The solar
abundances used are as given by \citet{gs98}.  Some observational data
based on the old solar abundances given by \citet{ag89} have been
corrected to be compatible with the new more recently determined solar
values.

\section{The basic galaxy formation model}
\label{galform}

We calculate galaxy formation in the framework of the CDM model using
the semi-analytical model {\sc galform} described in
\citet[][hereafter CLBF]{clbf00}, with some modifications as described
in \citet[][hereafter B04]{baugh04b}. The semi-analytical model
calculates various processes as follows: (a) dark matter halos form by
hierarchical merging; (b) gas falls into these halos and is
shock-heated; (c) gas which cools in halos collapses to form
rotationally supported galactic disks; (d) stars form {\em quiescently}
in disks; (e) supernova explosions inject energy into the gas,
ejecting some of it from the galaxy; (f) dynamical friction brings
galaxies together within common dark halos, leading to galaxy mergers;
(g) galaxy mergers can cause the transformation of stellar disks into
spheroids, and can also trigger {\em bursts} of star formation. The
model also computes the luminosities and colours of galaxies from
their star formation histories and metallicities, using a stellar
population model and including a treatment of dust extinction.

The CLBF model assumed a cosmic baryon fraction $\Omega_{\rm b}=0.02$,
that all stars formed with a Kennicutt (1983) IMF, and that gas
ejected from galaxies by supernova feedback was retained in the host
dark halo. \citet{bbflbc03} showed that if the more recent estimate
$\Omega_{\rm b}=0.04$ was used, the original model predicted too many
very luminous galaxies. They proposed two possible solutions to this
problem: either (i) {\em thermal conduction} suppresses the cooling of
gas in high mass halos, or (ii) gas is ejected from the halos of
galaxies by {\em superwinds}. B04 adopted the {\em superwind}
solution, but showed that in order to explain the numbers of
Lyman-break and sub-mm galaxies observed at high redshift, further
changes are required in the model: they proposed changes in the star
formation timescale in galaxy disks, in the triggering of starbursts
by galaxy mergers, and in the IMF of stars formed in bursts. In this
paper, we adopt the B04 version of our galaxy formation model as the
standard one, except that we retain both the {\em thermal conduction}
and {\em superwind} variants. Our implementation of these two
processes is similar to, but somewhat simpler than that of
\citet{bbflbc03}. For comparison purposes, we will also present some
results for the original unmodified CLBF model. In the remainder of
this section, we give some more details about some of the ingredients
of our galaxy formation model. Our modelling of chemical enrichment is
described in Section \ref{metals}.

{\em Halo mergers and gas cooling:} The merger histories of dark halos
are generated for {\it parent} halos at $z=0$, for a given power
spectrum of density fluctuations, using a Monte Carlo scheme based on
the extended Press-Schechter (PS) formalism (Press \& Schechter 1974;
CLBF).  The lowest mass progenitor dark halos are assigned baryons in
the form of diffuse gas only, according to the universal baryon
fraction $\Omega_{\rm b}/\Omega_{0}$.  The diffuse gas is raised by
shock heating to the virial temperature corresponding to the depth of
the halo potential well.  A fraction of the diffuse hot gas cools and
falls into the centres of the host dark halos, where it accretes onto
the disk of the central galaxy.  The rate at which hot gas cools
within a dark halo is computed by assuming a spherically symmetric gas
distribution and using metallicity-dependent cooling functions. The
predictions of this simple cooling model compare favourably with the
results of direct gas-dynamics simulations \citep{y02, h03}.

{\em Star formation in disks:} The rate at which stars form from the
cold gas in a galactic disk is given by $\psi=M_{\rm cold}/\tau_{*}$,
where $M_{\rm cold}$ is the mass of cold gas.  CLBF used a star
formation timescale that depended on the dynamical time of the
galaxy. Here, we adopt a prescription without any scaling with
dynamical time, following B04:
\begin{equation}
 \tau_{*}=\tau_{*}^{0}
\left({V_{\rm disk}}/{200\rm ~km~s^{-1}}\right)^{\alpha_{*}},
\label{eq:sfr}
\end{equation}
where $\tau_{*}^{0}$ and $\alpha_{*}$ are free parameters independent
of redshift, whose values are set by requiring that the model
reproduce the observed gas mass to luminosity ratios of present-day
disk galaxies.

We employ the supernova feedback model described by CLBF, in which
cold gas is reheated and ejected into the halo by supernovae
explosions according to the prescription:
\begin{equation}
 \dot{M}_{\rm reheat}=\beta\psi,
\end{equation}
where 
\begin{equation}
\beta=(V_{\rm disk}/V_{\rm hot})^{-\alpha_{\rm hot}}, 
\label{eq:SNfeedback}
\end{equation}
and $V_{\rm hot}$ and $\alpha_{\rm hot}$ are free parameters.  The
latter is fixed to $\alpha_{\rm hot}=2$ and $V_{\rm hot}$ is tabulated
in Table~\ref{tab:param} for the models considered in this paper.

As described above, for the standard baryon fraction employed in the
model, $\Omega_{b}=0.04$, our simple cooling and feedback models
result in the formation of too many giant luminous galaxies compared
with recent determinations of the galaxy luminosity function
\citep[e.g.][]{norberg02}.  We have investigated two different
approaches to solving this problem, which we now describe: (i) the
suppression of gas cooling by thermal conduction and (ii) the
suppression of star formation by galactic superwinds.

{\em Thermal conduction:} In the first approach, thermal conduction of
energy from the outer to the inner parts of the gas halo acts to
balance the energy removed from the gas by radiative cooling.  This
scenario is suggested by the lack of emission lines from low
temperature gas in the X-ray spectra of clusters
\citep[e.g.][]{peterson01}.  To model this phenomenon in a simple way,
we completely suppress cooling in halos whose circular velocity
exceeds $V_{\rm cond}$.  An estimate of the value of $V_{\rm cond}$
and its dependence on redshift can be derived by balancing the cooling
rate with the heating rate due to thermal conduction:
\begin{equation}
 n_{\rm e}n_{\rm i}\Lambda(T)\leq 
\nabla\cdot(\kappa\nabla T)\simeq \kappa(T) T/R^{2},
\end{equation}
where $n_{\rm e}$ and $n_{\rm i}$ are the number densities of
electrons and ions respectively, estimated here as their mean values
within the virial radius, $\Lambda(T)$ is the cooling function,
$\kappa(T)$ is the thermal conductivity, $T$ is the temperature of hot
gas, and $R$ is the radius of the halo \citep[e.g.,][]{tt81}. Assuming
a mean halo density given by the spherical collapse model, the Spitzer
form of conductivity without saturation, $\kappa\propto T^{5/2}$, and
a power law cooling rate for thermal bremsstrahlung, $\Lambda\propto
T^{1/2}$, we obtain
\begin{equation}
 V_{\rm cond}=V_{\rm cond}^{0}(1+z)^{3/4}.
\end{equation}
Here $V_{\rm cond}^{0}$ is given by
\begin{equation}
 V_{\rm cond}^{0}\simeq 200 \, \frac{f_{\rm hot}^{1/2}}{f_{\rm Sp}^{1/4}} 
\left( \frac{\Delta_V}{100} \right)^{1/4} {\rm  ~km~s}^{-1},
\end{equation}
where $f_{\rm hot}$ is the ratio of hot gas to total baryon mass in a
halo, $f_{\rm Sp}$ gives the ratio of the conductivity to the Spitzer
value, and $\Delta_V$ is the mean overdensity of a virialized halo
relative to the average density of the universe. (We have assumed
$\Lambda = 1.7\times 10^{-27} {\rm erg\, cm^3 s^{-1}} (T/{\rm
K})^{1/2}$ and $\kappa_{\rm Sp} = 1.0\times10^{-6}{\rm erg \, cm^{-1}
s^{-1} K^{-1}} (T/{\rm K})^{5/2}$.) For a flat universe with present
day density parameter $\Omega_{0}=0.3$, $\Delta_V \approx 300$
\citep{eke96}. Assuming $f_{\rm hot}=0.8$ (a realistic value in our
models) then gives $V_{\rm cond}^{0}=240{\rm kms}^{-1}$ for conduction
at the Spitzer rate ($f_{\rm Sp}=1$). To match the bright end of the
observed luminosity function, we show in Section 4 that we require
$V_{\rm cond}^{0}=100 {\rm kms}^{-1}$; this implies $f_{\rm Sp}\simeq
20$ if $f_{\rm hot}=0.8$.  While the cut-off circular velocity
required increases to $\sim 200$ km s$^{-1}$ if the redshift
dependence of $V_{\rm cond}$ is neglected, the implied efficiency of
conduction still seems unrealistically high.  

If conduction at this level is not viable, another mechanism needs to
be invoked in order to prevent too much gas from cooling in massive
halos.  One plausible, but as yet relatively unexplored alternative
mechanism is the heating of the hot gaseous halo by emission from an
active galactic nucleus \citep[e.g.][]{granato04}.  This may be
particularly relevant in the more modest mass halos in which the
merger rate peaks \citep{kh00, eng03}.  Here we do not enter into the
the details of the growth and fuelling of supermassive black holes in
galaxy mergers.  However, we retain the freedom to choose apparently
implausibly low values of $V_{\rm cond}$ on the grounds that other
phenomena in addition to conduction may be operating within the halo,
with the same consequences, namely the suppression of gas cooling in
halos above a certain mass.

{\em Superwinds:} In the second approach we have taken to prevent the
formation of too many bright galaxies, we assume that gas is ejected
out of galaxy halos by superwinds. The superwind is driven by star
formation, and ejects cold gas out of the galaxy and out of the halo
at a rate given by
\begin{equation}
 \dot{M}_{\rm SW}=\beta_{\rm SW} \,\psi,
\end{equation}
where 
\begin{equation}
\beta_{\rm SW} = f_{\rm SW} \, \min\left[1,(V_{\rm
    c}/200{\rm~km~s}^{-1})^{-2} \right], 
\label{eq:sw}
\end{equation}
$f_{\rm SW}$ is a parameter, and $V_{\rm c}$ is the circular velocity
of the galactic disk for quiescent star formation or the circular
velocity of the bulge in a starburst.  In practise, we adopt the same
values for the strength of the superwind, $f_{\rm SW}$, in the
quiescent and burst modes of star formation.  This is a simplified
version of the scheme proposed by \citet{bbflbc03}.  The form adopted
for $f_{\rm SW}$ means that the superwind becomes less effective at
ejecting cold gas from galaxies with $V_{\rm c}> 200{\rm kms}^{-1}$.
The gas (and associated metals) expelled from the halo are assumed to
be recaptured once the halo is subsumed into a more massive object
with a halo circular velocity in excess of $V_{\rm recap}$.  The free
parameters in the superwind feedback are thus $f_{\rm SW}$ and $V_{\rm
recap}$.

{\em Galaxy mergers and starbursts:} Following a merger of dark matter
halos, the most massive progenitor galaxy becomes the central galaxy
in the new halo, and the remaining galaxies become its satellites.
The orbits of the satellite galaxies decay due to dynamical friction.
If the dynamical friction timescale of a satellite is shorter than the
lifetime of the new halo (defined as the time taken for its mass to
double), the satellite will merge with the central galaxy.  The
consequences of the galaxy merger are determined by the mass ratio of
the satellite to the primary, $M_{\rm sat}/M_{\rm cen}$.  If $M_{\rm
sat}/M_{\rm cen} > f_{\rm ellip}$, the merger is classified as a {\em
major merger}; the merger remnant is a spheroid and any cold gas is
consumed in a starburst. The duration of the starburst is proportional
to the dynamical timescale of the bulge.  For mergers with $M_{\rm
sat}/M_{\rm cen} < f_{\rm ellip}$, the merger is classified as a {\em
minor merger}. In this case, the stars of the satellite are added to
the bulge of the primary, but the stellar disk of the primary remains
intact. If $M_{\rm sat}/M_{\rm cen} > f_{\rm burst}$ and the ratio of
cold gas mass to total stellar and gas mass exceeds $f_{\rm
gas,crit}$, the minor merger triggers a burst of star formation in
which all the gas from the primary disk participates.  We adopt the
same values for the parameters governing the physics of galaxies
mergers as used by B04: $f_{\rm ellip}=0.3, f_{\rm burst}=0.05$ and
$f_{\rm gas,crit}=0.75$.  We refer the reader to that paper for
further discussion of the merger model.

The standard model presented in this paper in both its {\em thermal
conduction} and {\em superwind} variants is able to successfully
reproduce many observed properties of present-day galaxies, such as
luminosity functions, gas fractions, and sizes of galaxy disks, as
will be shown  in \citet{l05}. The comparison of the {\em superwind}
variant of the model with observations of Lyman-break and sub-mm
galaxies at high redshift has already been published in
\citet{baugh04b}. The predictions for stellar metallicities of
galaxies will be presented in a subsequent paper \citep{n05}.

\begin{figure}
\epsfxsize=\hsize
\epsfbox{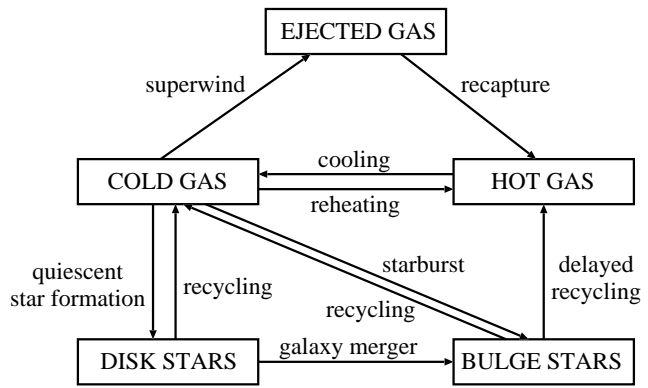}

\caption{Diagram showing the different baryonic components included in
the model, and the mechanisms by which mass and metals are exchanged
between them.}
\label{fig:exchanges}
\end{figure}

\section{STELLAR EVOLUTION AND METAL ENRICHMENT}
\label{metals}

Our galaxy formation model thus has baryons in four different
components: stars in galaxies, cold gas in galaxies, hot gas in halos,
and hot gas outside halos. There are exchanges of material between
stars and cold gas, cold gas and hot gas inside halos, cold gas and
hot gas outside halos, and hot gas outside and inside halos, as
described in the previous section. In addition, mergers of halos mix
their reservoirs of hot gas, and mergers of galaxies mix the stellar
and cold gas components of the two galaxies. The exchanges of mass and
metals between the different baryonic components are shown
schematically in Fig.~\ref{fig:exchanges}. Our model keeps track of
the effects of all of these mass exchanges and merging events on the
metallicities of the different components. Finally, we must calculate
the production of new heavy elements inside stars and their ejection
back into the gas phase. We describe now how we model this last
process.

\subsection{Evolution of a single stellar population}

Stars lose mass through stellar winds as they evolve. Some stars
ultimately undergo supernova explosions which pollute the interstellar
medium with newly produced metals.  The rates of both of these
phenomena depend upon the mass of the star.  The total amount of gas
recycled and metals produced per unit stellar mass depends, therefore,
on the adopted IMF.  In this Section we explain how we obtain the gas
restitution rate, the rates of SNe~II and SNe~Ia and the chemical
yield for a single stellar population.  This information is then 
incorporated directly into the {\sc galform} model, which predicts the
star formation histories of individual galaxies, in order to obtain
these quantities for the model galaxies.  The discussion below is
based upon that given by \citet{lpc02} with slight differences.
Stellar data are extracted from \citet{pcb98} for massive stars and
\citet{m01} for intermediate and low mass stars.  The chemical yields
of SNe~Ia are taken from \citet{n97}.

\begin{figure}
\epsfxsize=\hsize
\epsfbox{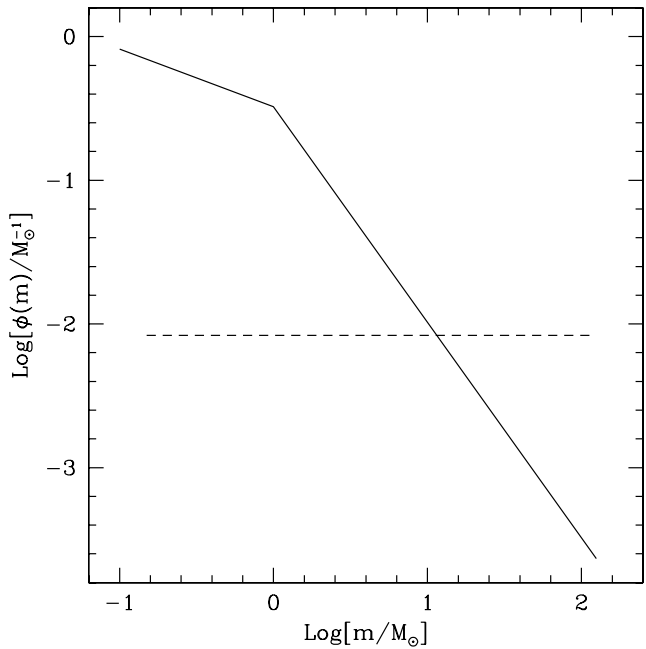}

\caption{The initial mass functions (IMFs) of star formation used in
this paper, $\phi(m)\equiv{\rm d}N/{\rm d}\ln m$.  For quiescent star
formation we use the Kennicutt (1998) IMF (solid line) and for
starbursts we use a flat IMF (dashed line). In both cases the IMFs are
normalised to be unity when integrated over the whole range of mass.
}
\label{fig:imf}
\end{figure}

We define the IMF as the number of stars per logarithmic interval of
stellar mass per unit total mass of stars formed:
\begin{equation}
\phi(m)\equiv{\rm d}N/{\rm d}\ln m \propto m^{-x}, 
\end{equation} 
normalised to unity when integrated over the whole range of mass
\begin{equation}
\int_{m_{d}}^{m_{u}} m \, \phi(m) \, d\ln m = 
\int_{m_{d}}^{m_{u}}\phi(m) \, dm = 1.
\end{equation} 
so that $\phi(m)$ has units of $M_{\odot}^{-1}$.  In this paper, we
use the IMF proposed by \citet{k83} for quiescent star formation in
disks, and a flat IMF in starbursts triggered by galaxy mergers.  The
\citet{k83} IMF has $x=0.4$ for $m< 1 M_{\odot}$ and $x=1.5$ for
$m\geq 1 M_{\odot}$. In starbursts, $x=0$.  In both modes of star
formation, the lower and upper stellar mass limits of the IMF are
$(m_{d},m_{u})=(0.15M_{\odot}, 120M_{\odot})$.  Fig.~\ref{fig:imf}
shows the quiescent (solid line) and starburst (dashed line) IMFs.

In addition to the visible stars with $m>0.15 M_{\odot}$, we allow for
brown dwarfs with $m<0.15 M_{\odot}$.  The mass fraction of brown
dwarfs is quantified by $\Upsilon$, which is the ratio of the total
mass in stars to the mass in visible stars for a newly formed
population of stars. Thus, the mass fraction in visible stars is
$1/\Upsilon$. CLBF adopted the Kennicutt IMF in both the quiescent and
burst modes of star formation, and required $\Upsilon=1.38$ to match
the predictions of their fiducial model to the observed present-day
galaxy luminosity function around $L_{*}$.  The other models we
consider in this paper have $\Upsilon=1$.

\begin{figure}
\epsfxsize=\hsize
\epsfbox{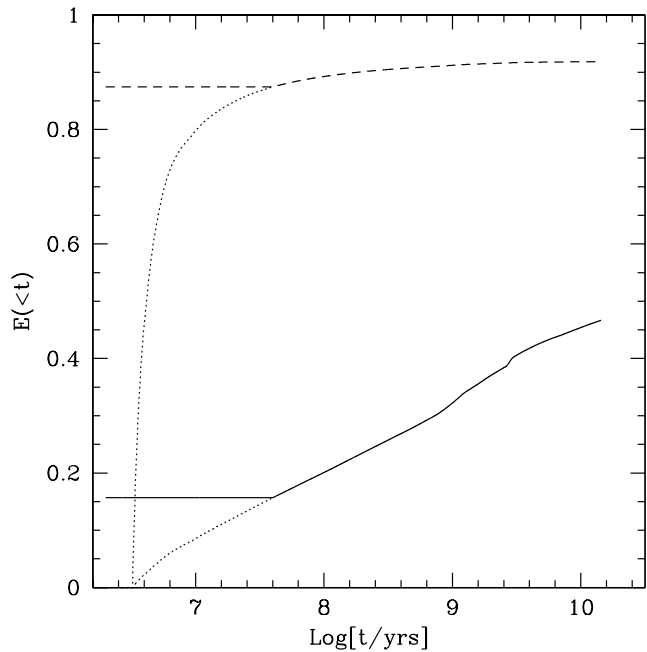}

\caption{The cumulative returned mass fraction, $E(\leq t)$, from
 evolved stars after time $t$ has elapsed from the epoch of star
 formation.  The solid and dashed lines indicate the $E(\leq t)$ we
 use for the Kennicutt and top-heavy IMFs, respectively.  Because we
 assume the instantaneous recycling for SNe~II, the values are
 constant until $\tau(8M_{\odot})$, where $\tau(m)$ is the lifetime of
 a star of mass $m$.  The dotted lines show how these results change
 when the instantaneous recycling approximation is dropped.}
\label{fig:e}
\end{figure}

The recycled mass fraction from an evolving population of stars is a
function of time.  For a population of stars formed with an IMF
$\phi(m)$ at time $t=0$, the total mass fraction released by time $t$
is
\begin{equation}
 E(\leq t)=\int_{M(t)}^{m_{u}}[m-M_{\rm r}(m)] \, \phi(m)\frac{dm}{m},
\label{eq:recycle}
\end{equation}
where $M(t)$ is the mass of a star reaching the end of its life at
time $t$, and $M_{\rm r}(m)$ is the mass of the remnant left by a star
with initial mass $m$.  In our model, since the timestep we use is
roughly comparable to the lifetime $4\times 10^7$yr of an 8$M_{\odot}$
star, the lowest mass star that results in a SNe~II, we can
justifiably use the instantaneous recycling approximation for SNe~II.
We therefore instantaneously release the mass recycled by stars with
$M\ge8M_{\odot}$.  In Fig.~\ref{fig:e}, we show $E(\leq t)$ for both
IMFs considered.  The dotted lines show how these predictions change
if the assumption of instantaneous recycling is relaxed.

The cumulative number of SNe~II explosions up to time $t$ (per
$M_{\odot}$ of stars formed), $R_{\rm II}(\leq t)$, and the mass
fraction of metals of the $i$-th element, $E_{i}(\leq t)$, released up
to time $t$ are given respectively by
\begin{eqnarray}
 R_{\rm II}(\leq
 t)&=&\int_{\max\{M(t),8M_{\odot}\}}^{m_{u}}\phi(m)\frac{dm}{m},
\label{eq:rateSNII}
\\
 E_{i}(\leq t)&=&\int_{M(t)}^{m_{u}}M_{i}(m)\phi(m)\frac{dm}{m},
\end{eqnarray}
where $M_{i}(m)$ is the mass of the $i$-th element released from a
star with mass $m$.  Because $E_{i}(\leq t)$ includes metals that
already existed when the stars formed, these pre-existing metals must
be subtracted in order to estimate the chemical yields, $p_{i}(\leq
t)$.  If the metallicity of stars with respect to the $i$-th element
at their birth is $Z_{i}^{0}$, $E_{i}$ can be divided into two terms,
\begin{equation}
 E_{i}(\leq t)=p_{i}(\leq t)+Z_{i}^{0}E(\leq t),
\end{equation}
where
\begin{equation}
 p_{i}(\leq t)=\int_{M(t)}^{m_{u}}y_{i}(m)\phi(m)\frac{dm}{m},
\label{eq:yieldSNII}
\end{equation}
and $y_{i}(m)$ indicates the mass of newly produced $i$-th element in a
star with mass $m$. Portinari et al. (1998) show that the dependence
of  $M_{i}$ and $y_{i}$ on initial stellar metallicity is small, so
for simplicity we neglect this dependence and use the values for solar
initial metallicity throughout.

\begin{figure}
\epsfxsize=\hsize 
\epsfbox{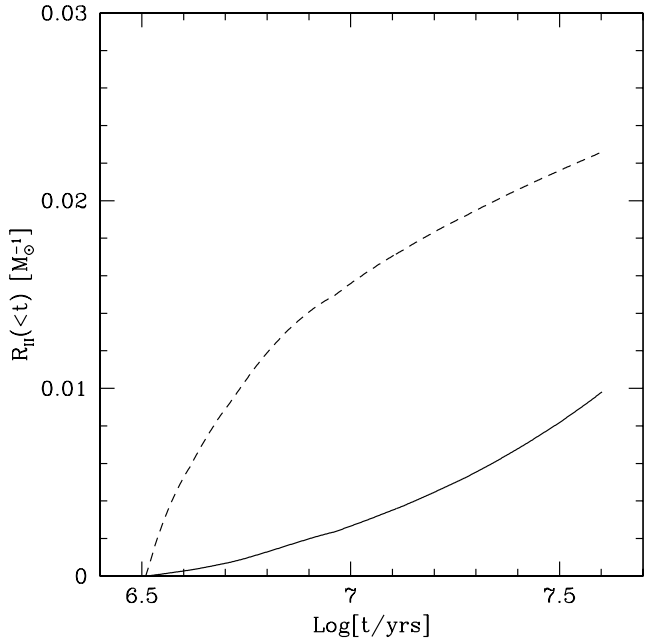}

\caption{The cumulative SNe~II rate.  
The lines types are the same as in  Fig.~\ref{fig:imf}.}
\label{fig:r2}
\end{figure}

\begin{figure}
\epsfxsize=\hsize
\epsfbox{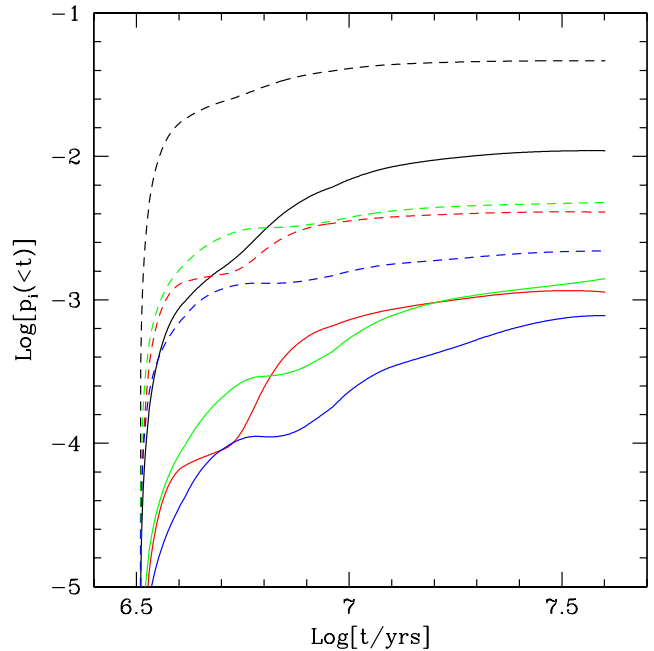}

\caption{The cumulative chemical yields of SNe~II, $p_{i}(\leq t)$.
 The solid and dashed lines denote the Kennicutt and top-heavy IMFs
 respectively.  For each case, the lines indicate O, Si, Mg and Fe in
 sequence from top to bottom at $t=\tau(8M_{\odot}) = 4\times
 10^7$yr.}

\label{fig:p}
\end{figure}

In Figs.~\ref{fig:r2} and \ref{fig:p}, we show $R_{\rm II}(\leq t)$
and $p_{i}(\leq t)$, respectively.  In both cases, the solid and
dashed lines denote the Kennicutt and top-heavy IMFs respectively.
The initial time for which results are plotted corresponds to
$\tau(m_{u})$ and the final time to $\tau(8M_{\odot})$, where
$\tau(m)$ is the lifetime of a star of mass $m$.  In Fig.~\ref{fig:p},
the lines indicate the yields of O, Si, Mg, Fe from top to bottom at
the right hand side of the panel.  In this paper, because
instantaneous recycling is assumed for SNe~II, we only use the values
of chemical yields at $\tau(8M_{\odot})$.

Note that we rescale the Mg yields from SNe~II, $p_{\rm Mg}$, upwards
by a factor of 4, so that the Mg abundances predicted by the model are
more consistent with observed abundances in solar
neighbourhood stars. This correction is standard practice in chemical
evolution modelling, and is justified because the theoretical
calculations of the yield of Mg are known to have a large (factor
$\sim 3$) uncertainty \citep{tww95, tgb98, pcb98, lpc02}. In reality,
any correction to the yields $y_{Mg}(m)$ is likely to depend on
stellar mass, so the correction factor for the integrated yield
$p_{\rm Mg}$ could be different for the two IMFs we use. In our model,
the enrichment of galactic disks is dominated by stars formed with the
Kennicutt IMF, while the $\alpha$-element enrichment of the ICM is
predicted to be dominated by stars formed with the top-heavy IMF, and
in fact, comparison with solar neighbourhood abundances favours a lower
correction factor ($\sim$2-3) for the Mg yield than do the ICM
abundances. However, for simplicity we have assumed the same
correction factor for both IMFs.


The standard scenario \citep{wi73} for SNe~Ia is that they 
occur in binary systems, in which a C-O white dwarf produced by
evolution of the primary star accretes gas from the secondary star 
when the secondary overflows its Roche lobe at the end of its own
evolution. This gas accretion drives the mass of the primary white
dwarf over the Chandrasekhar limit, causing it to explode. We model
this using the scheme of \citet{gr83}, with parameters updated
according to \citet{pcb98}. The progenitors of SNe~Ia are assumed to
be binary systems with initial masses in the range $m_{B,\rm low} <
m_B < m_{B,\rm up}$, where $m_B \equiv m_1 + m_2$, and $m_1$ and $m_2$ are
the initial masses of the primary and secondary ($m_2\leq m_1$)
respectively. We assume $m_{B,\rm up} = 2m_{1,\rm max}$, where
$m_{1,\rm max}$ is the largest single-star mass for which the endpoint
is a C-O white dwarf. The binary star systems which are the
progenitors of SNe~Ia are assumed to have an initial mass function
$A\phi(m_B)$, where $\phi(m)$ is the same as for the single-star
IMF. The distribution of mass fractions for the secondary,
$\mu=m_{2}/m_{B}$, is assumed to have the form (normalised over the
range $0<\mu<1/2$)
\begin{equation}
 f(\mu)=2^{1+\gamma}(1+\gamma)\mu^{\gamma} \qquad\qquad (0<\mu<1/2).
\end{equation}
For a single generation of stars formed at $t=0$, the cumulative
number of SNe~Ia explosions up to time $t$ is then given by
\begin{equation}
 R_{\rm Ia}(\leq t) = A\int_{m_{B,\rm low}}^{m_{B,\rm up}} 
\phi(m_{B}) \left[\int_{\mu_{\rm min}(t)}^{1/2} f(\mu) d\mu \right] 
\frac{dm_{B}}{m_{B}},
\label{eq:rateSNI}
\end{equation}
where the lower limit
\begin{equation}
\mu_{\rm min}(t) = \max\left[ \frac{M(t)}{m_{B}}, 
\frac{m_{B}-m_{B,{\rm up}}/2}{m_{B}}\right].
\end{equation}
is set by the conditions that the secondary has evolved off the main
sequence and that $m_1 = m_B - m_2 \leq m_{1,\rm max}$. Following
Portinari et al. (1998), we adopt $m_{B,\rm low}= 3M_{\odot}$,
$m_{B,\rm up}= 12M_{\odot}$ and $\gamma=2$. The factor $A$ is usually
chosen (for given $m_{B,\rm low}$ and $\gamma$) to reproduce the
observed ratio of rates of SNe~Ia to SNe~II in spiral
galaxies. Following Portinari et al., we adopt $A=0.07$. This
observational normalisation is derived for an IMF similar to
Kennicutt, but for simplicity we assume that the same normalisation
applies to the top-heavy IMF also.

\begin{figure}
\epsfxsize=\hsize
\epsfbox{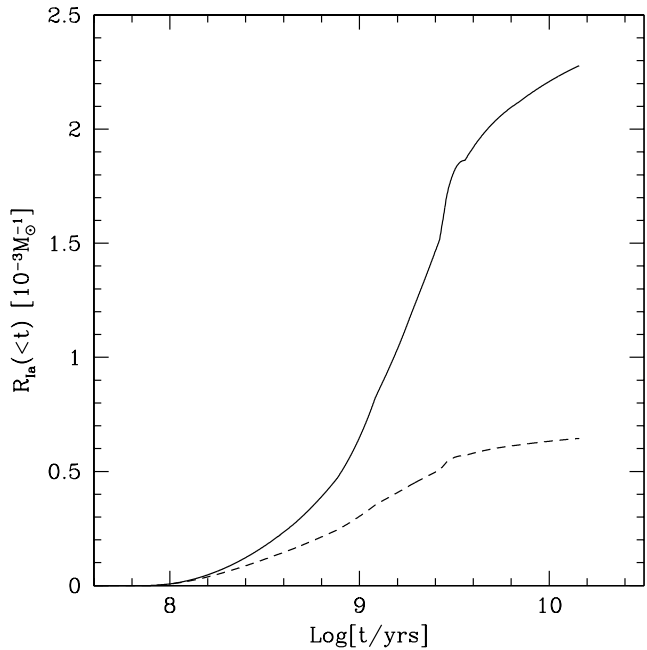}
\caption{The cumulative SNe~Ia rate, $R_{\rm Ia}$.  The line types are
the same as those used in Fig.~\ref{fig:imf}.  Note that the SNe~Ia
rate for the top-heavy IMF is smaller than that for the Kennicutt IMF,
because the top-heavy IMF has fewer intermediate mass stars.}
\label{fig:r1}
\end{figure}

In Fig.~\ref{fig:r1}, we show the cumulative number of SNe~Ia as a
function of time, $R_{\rm Ia}(\leq t)$, for the Kennicutt IMF (solid
line) and the top-heavy IMF (dashed line).  The number of intermediate
mass stars responsible for SNe~Ia is larger in the case of the
Kennicutt IMF than for the top-heavy IMF (as shown by
Fig.~\ref{fig:imf}).  This is in contrast to the situation for SNe~II.
The yield of metals from SNe~Ia is computed as
\begin{equation}
 p_{{\rm Ia},i}(\leq t)=M^{\rm Ia}_{i}R_{\rm Ia}(\leq t),
\label{eq:yieldSNI}
\end{equation}
where $M_{i}^{\rm Ia}$ is taken from the W7 model of \citet{n97} and
takes the values of 0.143, 8.5$\times 10^{-3}$, 0.154 and 0.626
$M_{\odot}$ for O, Mg, Si and Fe respectively.

Note that if the IMF includes a fraction of brown dwarfs (with
$\Upsilon > 1$), the values in equations (\ref{eq:recycle}),
(\ref{eq:rateSNII}), (\ref{eq:yieldSNII}), (\ref{eq:rateSNI}) and
(\ref{eq:yieldSNI}) must be multiplied by $1/\Upsilon$, because
$\Upsilon > 1$ corresponds to having a fraction $1/\Upsilon$ of
visible stars with $m_d<m<m_u$.

\subsection{The metal enrichment of cold and hot gas}
We treat the metal enrichment due to SNe~II in the same way as CLBF,
except that the instantaneously returned mass fraction and yields
(denoted $R$ and $p$ in CLBF) are replaced here by
$E(\leq\tau[8M_{\odot}])$ and $p_i(\leq\tau[8M_{\odot}])$. The mass
released from evolved stars after $\tau(8M_{\odot})$ and the metals
produced by SNe~Ia are computed at every timestep.  The gas and metals
recycled from disk stars are put into the cold gas in the galaxy,
while those from bulge stars are put into the hot gas in the halo.
This picture is consistent with many evolutionary models of disk
galaxies, in which metals from SNe~Ia are put into the gaseous disk;
observations suggest that elliptical galaxies do not keep the metals
released from SNe~Ia, because the metal abundances in the interstellar
medium of ellipticals are either smaller than or at best similar to
solar abundances \citep{mom00}.

The mass released from stars during the time interval
$t_{i} <t< t_{i+1}$ is computed from
\begin{equation}
 \Delta M=\int_{0}^{t_{i+1}}[E(\leq t_{i+1}-t')-E(\leq t_{i}-t')]\psi(t')dt',
\end{equation}
where $\psi(t)$ is the star formation rate at time $t$ for the galaxy
under consideration.  In order to solve this equation, the star
formation histories in both modes of star formation, quiescent and
starburst, must be recorded separately for individual galaxies.  Due
to the memory overhead involved, the star formation histories are
rebinned into 30 linear steps in time.  We have confirmed that our
results are insensitive to the precise number of steps used.

As well as the returned mass, the metals produced by evolving stars,
including SNe~Ia, are computed in a similar way.  The mass of the
$i^{\rm th}$ element released from evolved stars and SNe~Ia over the
time interval [$t_{i},t_{i+1}$] is
\begin{eqnarray}
\Delta M_{i}\hspace{-1mm}=\int_{0}^{t_{i+1}}M_{i}^{Ia}[R_{\rm Ia}(\leq
 t_{i+1}-t')-R_{\rm Ia}(\leq t_{i}-t')]\psi(t')dt'\nonumber\\
+\hspace{-1mm}\int_{0}^{t_{i+1}}\hspace{-1mm}Z(t')
[E(\leq t_{i+1}-t')-E(\leq t_{i}-t')]\psi(t')dt',
\end{eqnarray}
where $Z(t)$ is the metallicity of stars at their birth at time, $t$; 
this quantity is stored in the same fashion as the 
star formation histories.  The metals are also assumed to be put 
into cold and hot phases along with the recycled gas.

\section{The galaxy luminosity function}

In this paper we consider three principal models: the fiducial model
of CLBF (which assumed $\Omega_b=0.02$) and two new models which have
$\Omega_b=0.04$.  The first of these new models is {\em superwind}
model, in which gas ejection from the halo is invoked to suppress the
formation of bright galaxies.  In the second new model, which we refer
to as the {\em conduction} model, cooling in massive halos is
suppressed by thermal conduction. In the CLBF model, all stars form
with a Kennicutt IMF.  In the {\em superwind} and {\em conduction}
models, the Kennicutt IMF is adopted for quiescent star formation in
disks and a flat IMF is used in starbursts.  The superwind model is
similar to the model used by Baugh et al. (2004b).  We also consider a
variant of the superwind model (which we denote {\em
superwind/Kennicutt}) in which all star formation produces stars with
a Kennicutt IMF.  We list the parameters for the different models in
Table \ref{tab:param}.

\begin{table*}
\begin{center}
\caption
{ Model parameters.  Column~1 gives the name of the model, column~2
gives the baryon density parameter, column~3 gives $\Gamma$, the CDM
power spectrum shape parameter. Column~4 gives the dark matter halo
mass function used; J and PS denote mass functions given by
\citet{j01} and \citet{ps74} respectively. Columns~5 and 6 give the
parameters that specify the quiescent star formation timescale, as
defined by eqn.~(\ref{eq:sfr}).  Note that in the CLBF model, the star
formation timescale scales with the dynamical time of the disk,
$\tau_{\rm disk}$. Column~7 gives the value of the $V_{\rm hot}$
parameter that determines the strength of the standard SNe feedback
(eqn.~\ref{eq:SNfeedback}). Columns~8 and 9 describe the superwind
feedback; column~8 gives the parameter that sets the strength of the
superwind (eqn.~\ref{eq:sw}) and column~9 gives the threshold circular
velocity of halos above which gas ejected by a superwind can be
recaptured.  Column~10 gives the circular velocity of halos above
which cooling is suppressed in the conduction model. Column~11 gives
the IMF assumed in bursts; note that all models assume a Kennicutt IMF
for quiescent star formation. Finally, column~12 gives the ratio of
total stellar mass to the mass produced in luminous stars.
\label{tab:param}
}
\begin{tabular}{cccccccccccc}
\hline
(1) & (2) & (3) & (4) & (5) & (6) & (7) & (8) & (9) & (10) & (11) &
(12) \\
model & $\Omega_{\rm b}$ & $\Gamma$ & MF & 
 $\tau_{*}^{0}$ (Gyr) & $\alpha_{*}$ & $V_{\rm hot} ({\rm kms}^{-1})$& $f_{\rm SW}$ & $V_{\rm
 recap}$ (km~s$^{-1}$) & $V_{\rm cond}^{0}$ (km~s$^{-1}$) & burst IMF
& $\Upsilon$\\
\hline
 superwind  & 0.04 & 0.172  & J  & 8  & -3   & 200 & 2 & 600 & -- &
 x=0 & 1\\
 superwind/  & 0.04 & 0.172  & J  & 8  & -3   & 200 & 2 & 600 & -- &
 Kenn & 1\\
 \hfil Kennicutt &&&&&&&&&&&\\
 conduction & 0.04 & 0.172  & J  & 8   & -3  & 300 & 0 & --  & 100 &
 x=0 & 1\\
 CLBF       & 0.02 & 0.19   & PS & 200$\tau_{\rm disk}$ & -1.5 & 200 & 0
 & -- & --  & Kenn & 1.38\\
\hline
\end{tabular}
\end{center}
\end{table*}

\begin{figure}
\epsfxsize=\hsize
\epsfbox{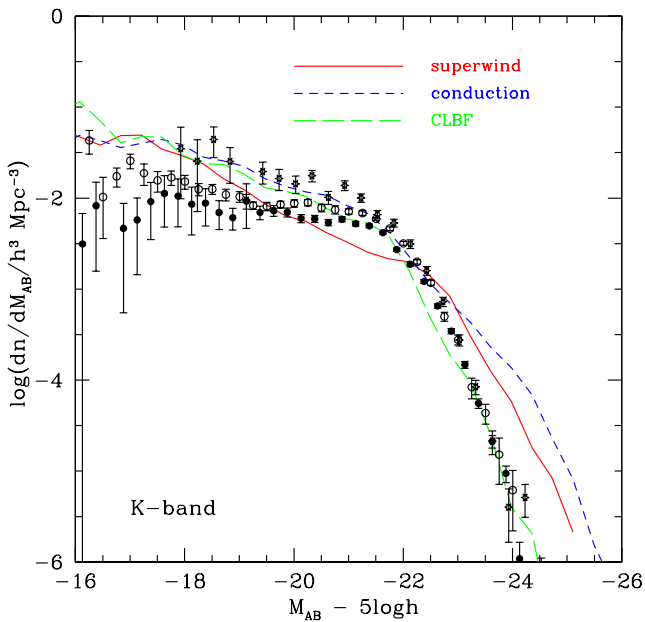}

\caption{The K-band galaxy luminosity function at $z=0$.  The model
predictions are shown by the lines as indicated by the key.  The
symbols with error bars show observational data from \citet[][ \it
filled circles]{c01}, \citet[][ \it stars]{hgct03}, and \citet[][ \it
open circles]{bmkw03}.}

\label{fig:lf}
\end{figure}

An important constraint on the models is that they should be
consistent with the observed present-day galaxy luminosity function.
In Fig.~\ref{fig:lf} we compare the model predictions for the
present-day $K$-band luminosity function with observational estimates.
The predictions of the superwind model with $V_{\rm recap}=600$
km~s$^{-1}$ are shown by the solid line, those of the conduction model
with $V_{\rm cond}^{0}=100$ km~s$^{-1}$ by the short dashed line, and
the luminosity function of the CLBF model is plotted using a long
dashed line.  The symbols show observational data. Although it is
challenging for models with the currently favoured value of
$\Omega_{\rm b}$ to match the observed luminosity function as well as
in the CLBF model, as shown by \citet{bbflbc03}, the superwind and
conduction models are in broad agreement with the observations.

In the conduction model, the value of the parameter $V_{\rm cond}^{0}$
which is required to produce a good match with the observed luminosity
function seems unphysically small.  The value of $V_{\rm
cond}^{0}=100{\rm kms}^{-1}$ requires a thermal conductivity that is
many times larger than the Spitzer value.  As explained in Section 2,
processes other than thermal conduction could also contribute to the
suppression of cooling, so we retain the freedom to adopt such a low
value for this parameter.
 
\begin{figure}
\epsfxsize=\hsize
\epsfbox{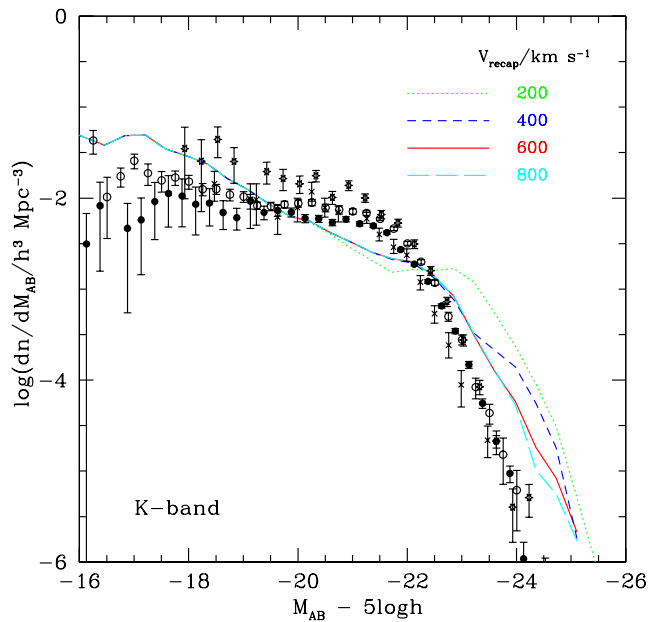}

\caption{The K-band luminosity function predicted in the superwind
model for different values of $V_{\rm recap}$, the threshold halo
circular velocity for the recapture of gas previously ejected by a
superwind (as indicated by the key).  The fiducial superwind model has
$V_{\rm recap}= 600{\rm kms}^{-1}$ and is shown by the solid line. The
observational data are plotted using the same symbols as in
Fig.~\ref{fig:lf}. }

\label{fig:lfVrecap}
\end{figure}

In Fig.~\ref{fig:lfVrecap} we show how the luminosity function in the
superwind model depends on the parameter $V_{\rm recap}$, the
threshold circular velocity above which halos recapture gas previously
ejected in a superwind. Clearly if $V_{\rm recap}$ is too low, then
too many very bright galaxies are produced. On the other hand, the
galaxy luminosity function becomes insensitive to increases of this
parameter above $V_{\rm recap} \approx 600$ km~s$^{-1}$. However, as we
will show later, the value of $V_{\rm recap}$ has a strong impact on
metal abundances in the ICM. The predicted ICM metal abundances are
too low unless the metals ejected in the superwind can be recaptured
by cluster-sized halos.  Thus, we require a value $V_{\rm recap}\simeq
600$ km~s$^{-1}$ in order to reproduce both the observed field galaxy
luminosity function and the metal abundance of the ICM.

\section{Present-day intracluster medium}
\subsection{Gas mass-to-stellar luminosity ratio and hot gas fraction}

A basic test for any model of the chemical enrichment of the ICM is that
it should reproduce the observed ratios of the ICM mass to total stellar
luminosity and to total mass in clusters. This is because the metals are
produced in stars, so a model which produces the observed ICM
metallicity is invalid unless it predicts the correct number of stars at
the same time.

\begin{figure}
\epsfxsize=\hsize
\epsfbox{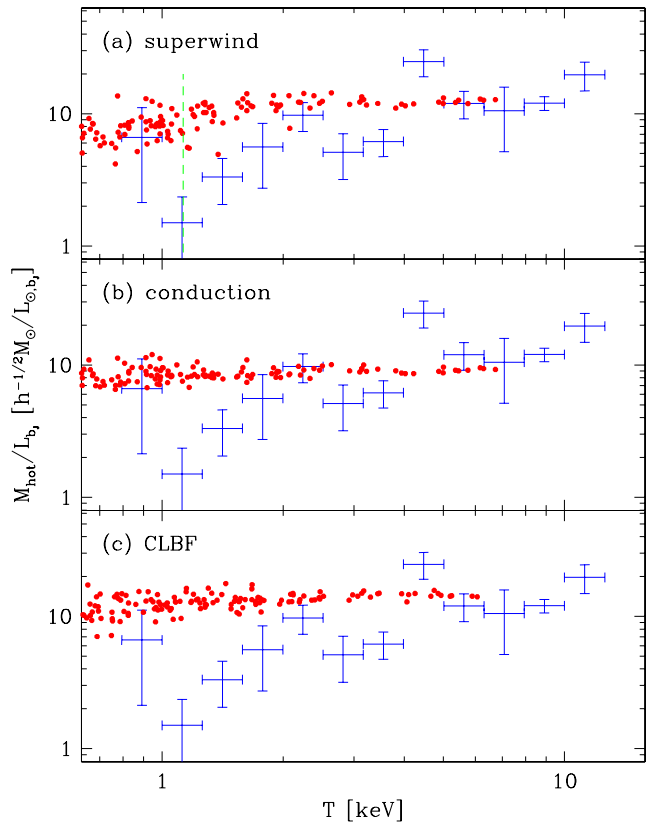}

\caption{The ratio of ICM mass to $b_{\rm J}$-band luminosity of stars
 contained in a cluster, $M_{\rm hot}/L_{b_{\rm J}}$, plotted as
 function of the hot gas temperature.  The dots show the predictions
 of: (a) the superwind model, (b) the conduction model, and (c) the
 CLBF model.  The crosses in all panels denote the binned averages of
 $M_{\rm hot}/L_{b_{\rm J}}$ estimated using observational data from
 \citet{spflm03} and \citet{sp03}.  The dashed vertical line in panel
 (a) shows the temperature corresponding to a halo circular velocity
 $V_{\rm recap}=600$ km~s$^{-1}$, above which ejected gas is
 recaptured.}

\label{fig:ML}
\end{figure}

In Fig.~\ref{fig:ML}, we plot the ratio $M_{\rm hot}/L_{b_{\rm J}}$ of
the mass of the ICM, $M_{\rm hot}$, to the total $b_{\rm J}$-band
stellar luminosity of the galaxies, $L_{b_{\rm J}}$, as a function of
the ICM temperature $T$. We show predictions for the superwind,
conduction and CLBF models. In the models, the ICM temperature is
assumed to equal the halo virial temperature,
\begin{equation}
T_{vir} = \frac{1}{2} \frac{\mu m_H}{k_B} V_c^2 = 3.1{\rm keV} 
\left( \frac{V_c}{1000 {\rm kms}^{-1}} \right)^2,
\end{equation}
while for the observations we use the measured X-ray temperature. The
relationship between cluster mass and virial temperature in the
$\Lambda$CDM model at $z=0$ is then $T_{vir} = 6.4{\rm keV}\,
(M/10^{15}h^{-1}M_{\odot})^{2/3}$.  We compare the models with
observational data from \citet{spflm03} and \citet{sp03}. We have
combined the data from individual clusters in bins in X-ray
temperature, and computed the mean and dispersion in each bin,
weighted according to the errors on the individual datapoints.  The
error bars on the plotted datapoints are calculated from the
dispersion in each bin, and do not include any other measurement
errors.

The conduction and CLBF models predict a ratio $M_{\rm hot}/L_{b_{\rm
J}}$ which increases only slightly with cluster temperature. The
superwind model predicts a somewhat larger increase, because of the
expulsion of gas from lower-mass halos, but again is almost flat for
halos with $V_c>V_{\rm recap}$, corresponding to $T\approx 1.1$keV which
is indicated by the vertical dashed line in the top panel. The models
are in approximate agreement with the observational data for $T\gtrsim
2$keV, but for lower temperatures, the observations suggest ICM
fractions significantly smaller than what the models predict. However,
there are still important uncertainties in the observational data (not
included in the error bars plotted) resulting from the need to
extrapolate X-ray measurements out to the cluster virial radius, so the
ICM fractions in cooler clusters are not yet securely determined as
shown below.

We note that the CLBF model predicts very similar $M_{\rm
hot}/L_{b_{\rm J}}$ values to the superwind and conduction models,
even though it has a global baryon fraction only half that of the
other models. This agreement comes about because the CLBF model
predicts a luminosity density which is a factor $\sim 2$ lower in all
environments compared to the other models.

\begin{figure}
\epsfxsize=\hsize
\epsfbox{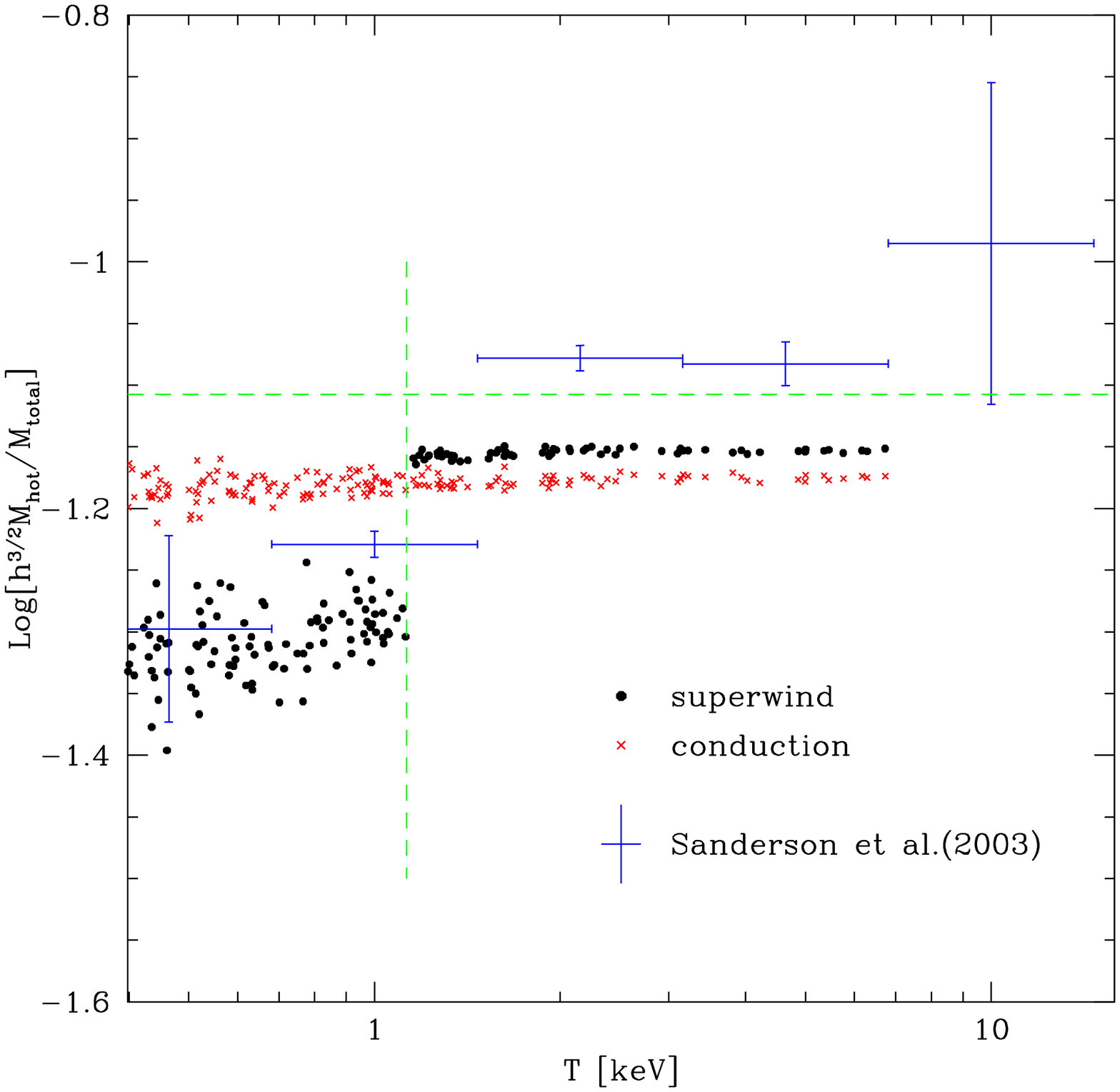}

\caption{The ratio of hot gas mass to total mass, including dark
matter, as a function of the hot gas temperature.  The dots and
crosses show the predictions of the superwind model and the conduction
model, respectively. The points with error bars show the binned
averages of the observational data from \citet{spflm03} (we use their
estimates of gas fraction wihthin the virial radius). The horizontal
dashed line indicates the universal ratio of baryon to total mass
density estimated from $\Omega_{0}$ and $\Omega_{\rm b}$. The vertical
dashed line shows the temperature corresponding to a halo circular
velocity $V_{\rm recap}=600$ km~s$^{-1}$, above which ejected gas in
the superwind model is recaptured.}

\label{fig:MgasMhalo}
\end{figure}

In Fig.~\ref{fig:MgasMhalo} we show the ratio of the ICM mass to the
total cluster mass including dark matter as a function of cluster
temperature. The dots and crosses indicate the hot gas fractions for
the superwind and the conduction models, respectively.  The dashed
horizontal line corresponds to the cosmic baryon-to-total mass ratio;
the model points would all lie on this line if all baryons were
retained in the cluster in the hot gas component. We also show
observational data from \citet{spflm03}, binned in temperature as in
Fig.~\ref{fig:ML}. The superwind model reproduces the qualitative
trend seen in the observations that the gas fraction drops in low
temperature clusters. However, we caution that the observational
estimates rely on extrapolating both the ICM and dynamical masses from
the outermost measured X-ray datapoint out to the halo virial
radius. Future more accurate X-ray observations will allow more
precise tests of models for gas ejection from halos.

\subsection{Metal abundances}
We now compare the ICM metallicities predicted by our models with
observational data. Our model predicts global metallicities, averaged
over the entire ICM within the cluster virial radius. In contrast, the
ICM metallicities measured from X-ray data are often only
representative of the central regions of the cluster, because these
dominate the X-ray emission.  Some clusters, known as {\it cooling
flow} clusters, show metallicity gradients for some elements, in the
sense that metallicities increase towards the cluster centre. It is
therefore essential to correct the observational data to estimate mean
metallicities averaged over the entire cluster, before comparing with
the models. We compare our model predictions with the following
observational data on ICM abundances: \citet[][ {\it BeppoSAX}]{dG03}
for Fe [see also \citet{ib01} and \citet{dG02}]; \citet[][ {\it
XMM-Newton}]{p03} for O, Fe, Mg and Si [see also \citet{tkhbp04}];
\citet[][ {\it ASCA}]{f98} for Fe and Si; and \citet[][ {\it
ASCA}]{blhm03} for Si. We explain in the Appendix why we have omitted
some observational data from our comparison. We use the results of the
systematic study of iron gradients by \citet{dG03} to convert the
measured central metallicities of iron and silicon to global average
values, assuming that silicon has the same gradient as iron, as is
suggested by observational data. We present the details of how we make
these corrections in the Appendix. The size of these estimated
corrections is up to a factor of two.  Oxygen and magnesium do not
show metallicity gradients \citep{tbkfm01, tkhbp04}, so we assume that
the global abundances of these elements are equal to the measured
central values. We note that previous studies (e.g. de Lucia et
al. 2004) have not included any correction for metallicity gradients
when comparing models with observed ICM abundances.

\begin{center}
\begin{figure*}
\epsfxsize=0.7\hsize
\leavevmode
\epsffile{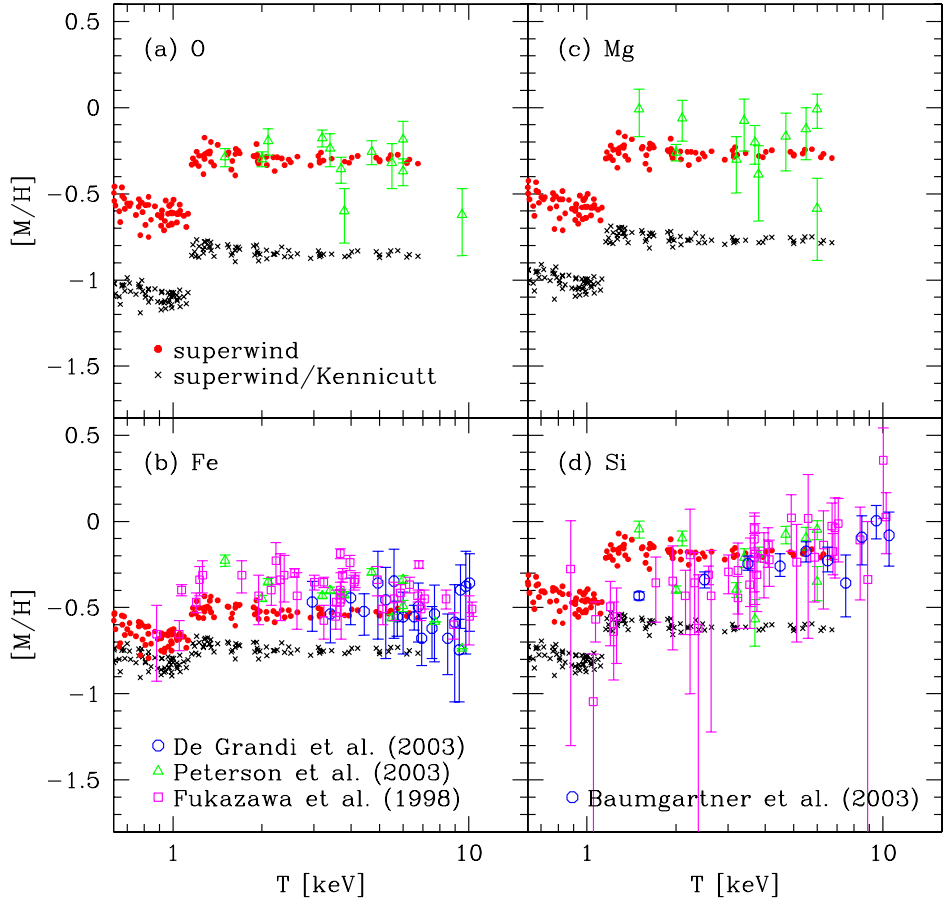}
\caption{ The metal abundances of the ICM.  Each panel shows a
different element: (a) [O/H] (b) [Fe/H] (c) [Mg/H] (d) [Si/H].
Abundances are normalized to solar values, and shown on a logarithmic
scale. The predictions of the superwind model are shown by dots in the
standard case where starbursts have a top-heavy IMF, and by crosses
for the variant model in which all star formation takes place with a
Kennicutt IMF.  The open triangles in each panel denote observational
data from {\it XMM-Newton} \citep{p03}, and the open squares data from
{\it ASCA} \citep{f98}; in panel (b), the open circles show data from
{\it BeppoSAX} \citep{dG03}, while in panel (d), the open circles show
{\it ASCA} data from \citet{blhm03}. All of the datapoints are
measurements for individual clusters, apart from \citet{blhm03}, which
are average values in bins of temperature. The observational data have
been corrected for abundance gradients as described in the text. }

\label{fig:MH}
\end{figure*}
\end{center}

We first show how the predicted metal abundances in the ICM depend on
the IMF adopted for starbursts.  Fig.~\ref{fig:MH} shows the O, Fe, Mg
and Si abundances (relative to solar values) as functions of cluster
temperature for the superwind model.  The dots show the predicted
values for the standard superwind model with a top-heavy IMF ($x=0$)
in bursts and a Kennicutt IMF in disks, and the crosses show a variant
superwind model in which disks and bursts both have a Kennicutt
IMF. For clusters with $T\gtrsim 1$keV, the abundances of the
$\alpha$-elements O, Mg and Si are 3--4 times higher in the case of
the top-heavy IMF in bursts, while the abundance of Fe is only about
60\% higher. This difference reflects that fact that the
$\alpha$-elements are produced mostly by SNe~II, and the number of
SNe~II is larger for a top-heavy IMF, while Fe is produced by both
SNe~Ia and SNe~II, and the number of SNe~Ia is actually reduced for
the top-heavy IMF. In fact, the models predict that most Fe is
produced by SNe~Ia for a Kennicutt IMF, but by SNe~II for the
top-heavy IMF.

When in Fig.~\ref{fig:MH} we examine the dependence of ICM abundances
on cluster temperature (or mass) in the superwind model, we see that
the abundances of the different elements all show a break at $T\approx
1.1$ keV. This temperature corresponds exactly to the halo circular
velocity $V_{\rm recap}=600$ km~s$^{-1}$ above which gas and metals
ejected by superwinds are recaptured. Halos with $V_c > V_{\rm recap}$
effectively trap all of the metals ever produced by the galaxies they
contain. For clusters hotter than the break, the abundances are
predicted to be almost constant. The jump in abundances across the
break is about a factor 2 for O, Mg and Si, but is much smaller for
Fe, especially if a Kennicutt IMF is used in bursts. This again
reflects the fractions of these different elements produced in either
SNe~II or SNe~Ia. Since SNe~II also drive the superwinds, while SNe~Ia
are typically delayed until after the strong superwind phase has
subsided, superwinds are much more effective in expelling from halos
the heavy elements produced in SNe~II than those produced in SNe~Ia.
If $V_{\rm recap}$ is varied, then the metal abundances in clusters
with $V_c>V_{\rm recap}$ are insensitive to the value of $V_{\rm
recap}$ for $V_{\rm recap}\ga 400$ km~s$^{-1}$, but increase slightly
if $V_{\rm recap} \la 400$ km~s$^{-1}$, because more stars are
produced in that case (c.f. Fig.\ref{fig:lfVrecap}).

Fig.~\ref{fig:MH} also shows the observed ICM abundances of O, Mg, Si
and Fe, corrected to global average values as we have described. We
see that the superwind model with a top-heavy IMF in bursts agrees
well with the observed abundances of O, Mg and Si, while the superwind
model with a Kennicutt IMF predicts abundances of these elements which
are too low by factors 2--3. For Fe, the model with the top-heavy IMF
again agrees better with the observations, although the abundance is
still slightly low. We note that the current abundance measurements
for O, Mg and Si are all for clusters hotter than the break predicted
in the models. However, the value of $V_{\rm recap}$ cannot be
significantly increased above our standard value of 600 km~s$^{-1}$
without moving the abundance break into the temperature range where
the observations indicate a plateau in abundances, thus conflicting
with the data. The measured Fe abundances, which extend down to
somewhat lower cluster temperatures, do in fact suggest a break close
to the temperature predicted by the superwind model, but more
measurements for cool ($T<1$keV) clusters are needed to confirm
this. It is clear that more abundance measurements in cool clusters,
especially of O, Mg and Si, would put stringent constraints on our
model of ejection of gas and metals from halos by superwinds. The
models also predict a very small scatter in global ICM abundances at a
given cluster temperature, especially above the break. The observed
abundances show a larger scatter, but most of this results from
measurement errors.

\begin{figure*}
\epsfxsize=0.7\hsize
\leavevmode
\epsffile{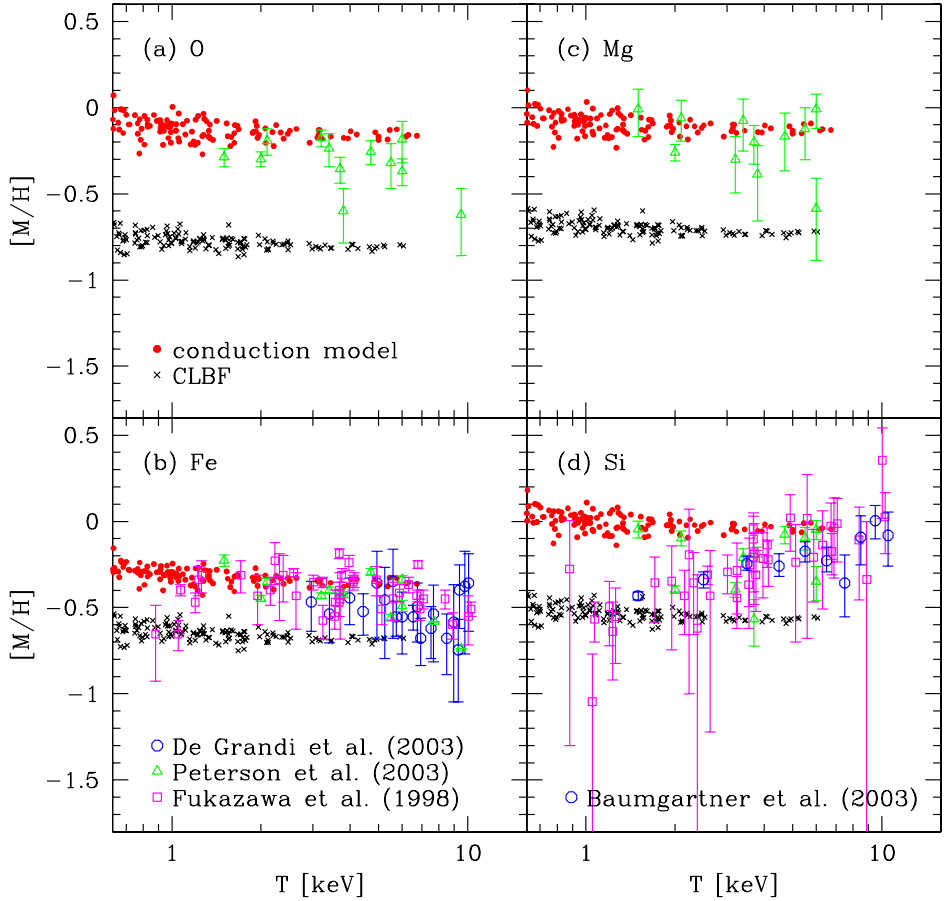}
\caption{The metal abundances in the ICM for the conduction model
(dots) and the CLBF model (crosses). The observational data are
plotted using the same symbols as in Fig.~\ref{fig:MH}.  }
\label{fig:MH2}
\end{figure*}

In Fig.~\ref{fig:MH2} we show the ICM metal abundances predicted by
our other two models: the conduction model with a top-heavy IMF in
bursts, and the CLBF model with a Kennicutt IMF for all modes of star
formation. The models are compared with the same observational data as
in Fig.~\ref{fig:MH}. For both of these models, halos retain all of
their metals, and consequently the predicted ICM abundances are almost
constant with cluster temperature. The CLBF model predicts abundances
too low compared to observed values, by factors 2--3 for O, Mg and Si,
and by around 40\% for Fe, and is thus ruled out. The conduction model
predicts abundances in much better agreement with observations,
although slightly too high for O and Si. The abundances in the
conduction model are slightly higher overall than those in the
superwind model.

It should be remembered that we have applied a correction to the yield
of Mg from SNe~II predicted by stellar models, in order to better
match observational constraints. The model yields have been multiplied
by a factor of four throughout, a value which is a compromise between
that required to reproduce Mg abundances in galactic disks
(specifically, in the solar neighbourhood) and in the ICM; a separate
correction factor could be applied to each IMF because of the
mass-dependent uncertainties in estimating theoretical yields.  No
correction factors have been applied to the yields of the other
elements: O, Fe and Si.

\subsection{Abundance ratios}
\begin{figure}
\epsfxsize=\hsize
\epsfbox{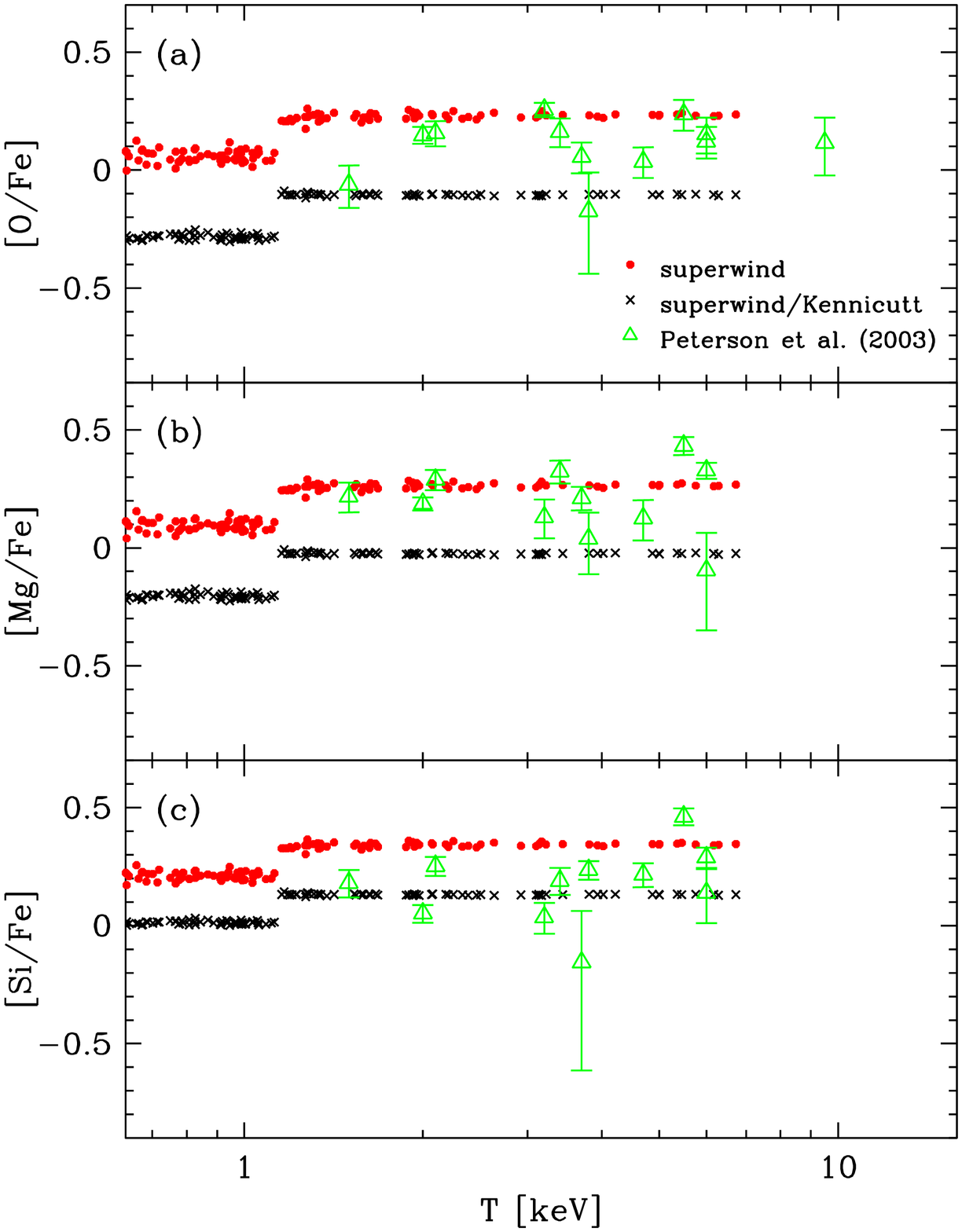}

\caption{The abundance ratios of $\alpha$-elements to iron. Each panel
corresponds to a different ratio: (a) [O/Fe].  (b) [Mg/Fe]. (c)
[Si/Fe]. All ratios are relative to solar values. The predictions of
the superwind model are plotted using dots for the standard case in
which starbursts have a top-heavy IMF, and crosses when all star
formation is with the Kennicutt IMF.  The open triangles show
observational data from \citet{p03}, corrected for radial gradients.}

\label{fig:MFe}
\end{figure}

Examining the abundance ratios between different elements in the ICM
provides an alternative method to constrain the form of the IMF, in
which the effects of uncertainties in the ICM to stellar mass ratio
cancel out.  In Fig.~\ref{fig:MFe} we show the abundance ratios of
different $\alpha$ elements to iron, [O/Fe], [Mg/Fe] and [Si/Fe], for
the standard superwind model, with a top-heavy IMF in bursts, and the
variant superwind model, with a Kennicutt IMF in bursts. We also show
observational data from \citet{p03}, which in the case of Si and Fe
have been corrected for abundance gradients in the way already
described. We see that the predicted O/Fe and Mg/Fe ratios agree
better with observations for the model with the top-heavy IMF, while
on the other hand the Si/Fe ratio agrees better with the model having
a Kennicutt IMF in bursts. The overall level of agreement of the
abundance ratios with observational data in the top-heavy IMF model
could be improved if the Fe abundances in the model were slightly
larger (this can also be seen in Fig.~\ref{fig:MH}), or if the
observed Si abundances were slightly larger. The latter would occur if
the abundance gradients in Si were weaker than we have assumed, since
we would then infer larger global Si abundances from the same measured
central values. However, recent data from \citet{tkhbp04}, in which
metal abundances are measured out to $r\la 500 h^{-1}$ kpc from
cluster centres, seem to support our original assumption that the
radial gradients in Si and Fe are the same.

\begin{figure}
\epsfxsize=\hsize
\epsfbox{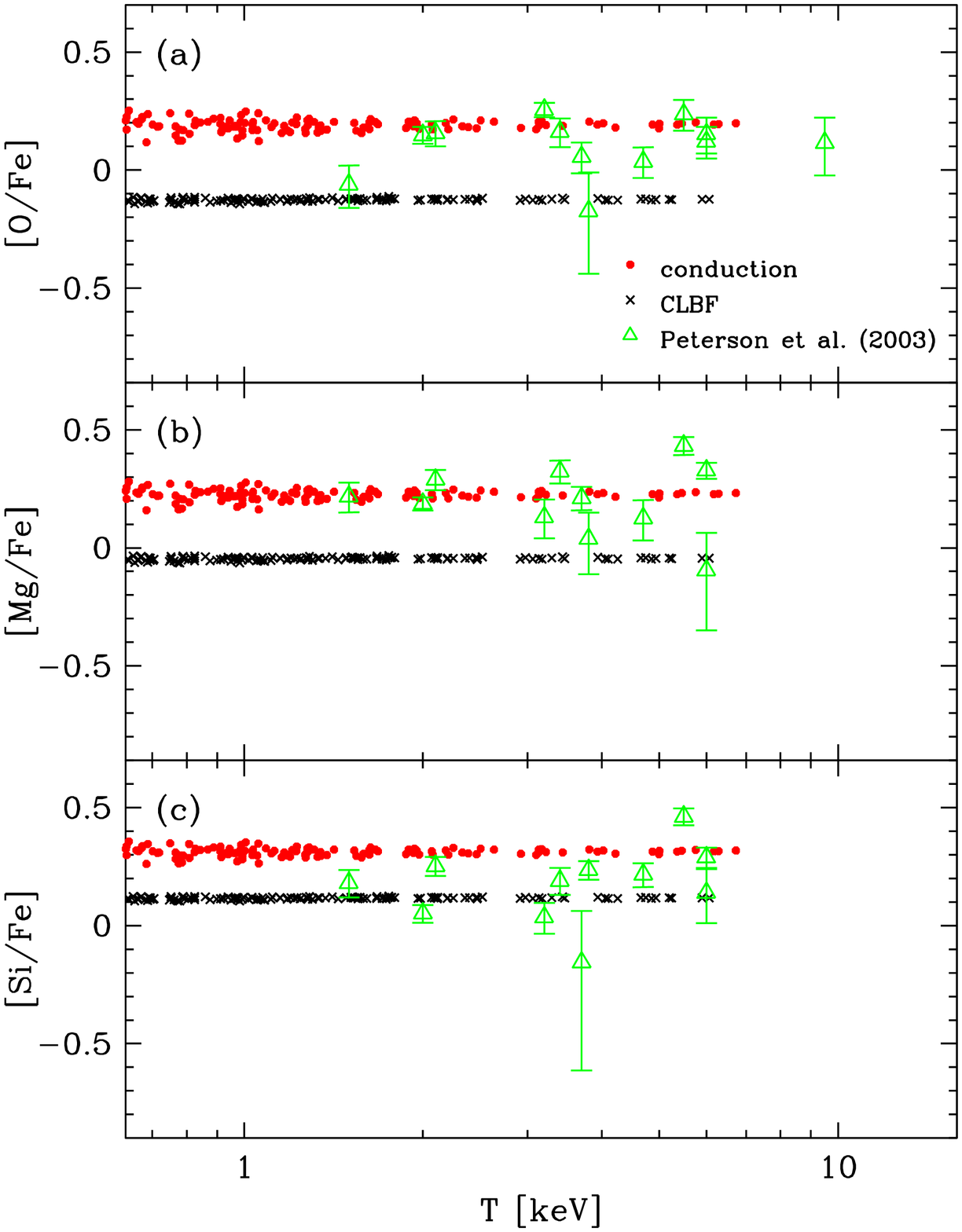}
\caption{ The abundance ratios in the conduction model (dots) and the
CLBF model (crosses). The data are the same as plotted in
Fig.~\ref{fig:MFe}.  }
\label{fig:MFe2}
\end{figure}

Fig.~\ref{fig:MFe2} is the equivalent of Fig~\ref{fig:MFe} for the
conduction and CLBF models.  Since the conduction model predicts 
slightly higher Fe abundances than the superwind model, the agreement
with the observed abundance ratios improves.  The CLBF model fails to
reproduce the observed abundance ratios. The models predict very
little variation in the abundance ratios with cluster temperature,
apart from the break in the superwind model where recapture sets
in. They also predict very small scatter for the abundance ratios at a
given cluster temperature.

\subsection{Metallicities of stars and gas in clusters}
\begin{figure}
\epsfxsize=\hsize
\epsfbox{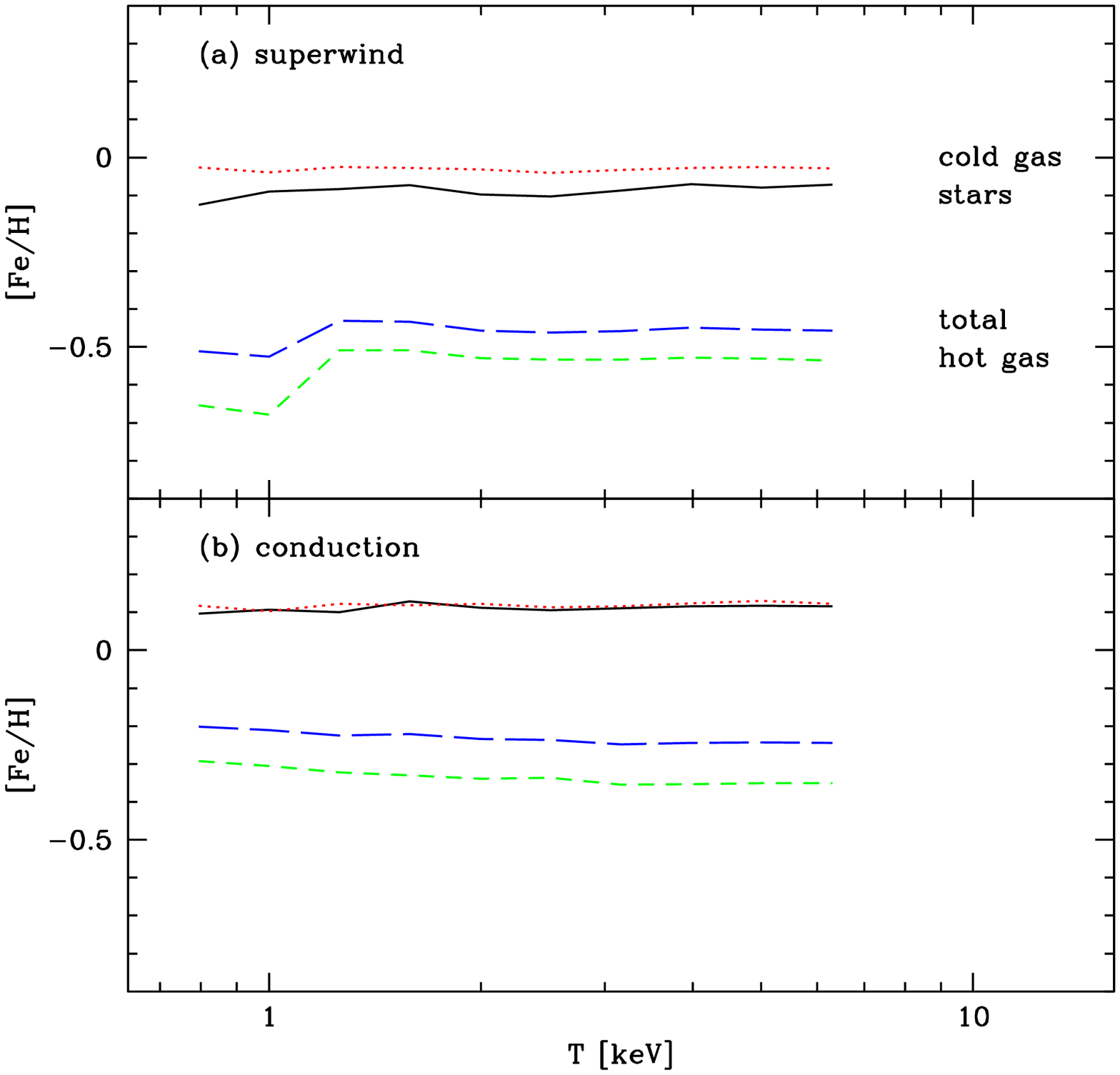}

\caption{The iron metallicities, [Fe/H], of stars, cold gas and hot
 gas, shown by the solid, dotted and short-dashed lines, respectively.
 The mass-weighted, mean total metallicities are shown by the
 long-dashed line. Panel (a) shows results for the superwind model and
 (b) for the conduction model.}

\label{fig:metalcomp}
\end{figure}

In Fig.~\ref{fig:metalcomp} we compare the predicted mass-weighted
mean iron metallicities for the different baryonic components in
clusters: stars in galaxies, cold gas in galaxies, and hot
intracluster gas. Results are shown for the standard superwind and
conduction models as functions of cluster temperature. In both models,
the mean stellar and cold gas metallicities are close to solar, while
the hot gas metallicity is 2--3 times lower. The mean total
metallicity is only slightly larger than the hot gas metallicity,
reflecting the fact that most of the baryons are in the hot gas
component. For the superwind model, the onset of recapture of gas by
halos produces the expected jump in the hot gas and total
metallicities, but does not produce any corresponding feature in the
metallicities of the stars or cold gas. This is because the enrichment
of these components has almost finished by the formation epoch of the
cluster and the epoch of recapture of expelled metals (see Section 7).

\begin{figure}
\epsfxsize=\hsize
\epsfbox{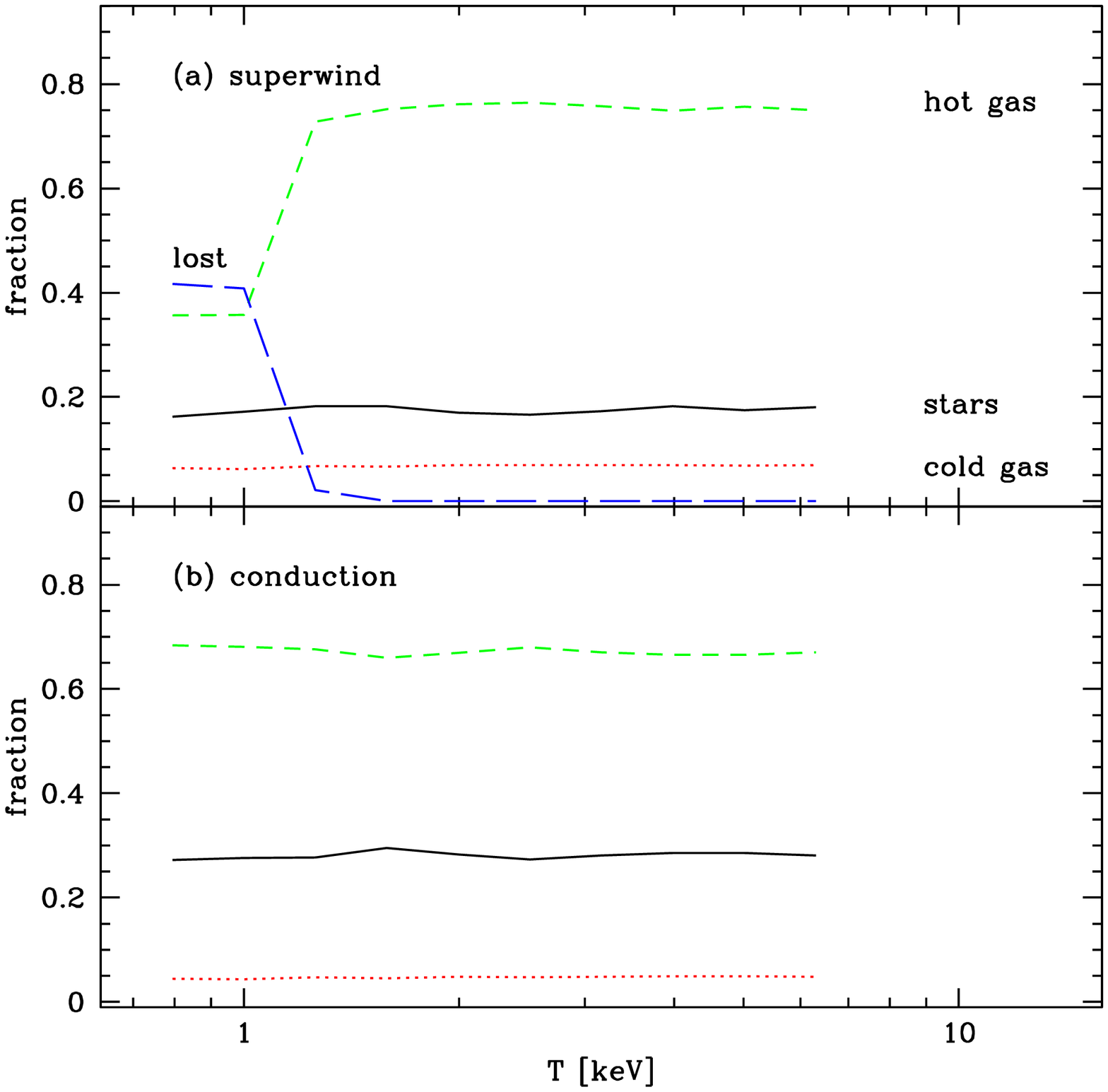}

\caption{The mass fractions of iron in different components.  The line
 types are the same as in Fig.~\ref{fig:metalcomp} except for the
 long-dashed line, which indicate the fraction of iron in gas
 expelled from halos (labelled as `lost').}

\label{fig:metalfrac}
\end{figure}

Fig.~\ref{fig:metalfrac} shows the fractions of the total mass of iron
which are in stars, cold gas, hot gas and `lost' gas which is expelled
from halos by superwinds, for the same models as in
Fig.~\ref{fig:metalcomp}. In all cases, more of the iron is in the hot
gas than in galaxies. In the superwind model, superwinds put about
40\% of the iron outside halos for low mass clusters below the
threshold for recapture. Thus, according to this model, a significant
mass of metals should be observed in the surroundings of low
temperature clusters.

Thus, in summary, from the ICM results at $z=0$, we conclude that a
top-heavy IMF with $x\simeq 0$ in starbursts is required for the model
predictions to match the observed ICM metallicities, as long as we
adopt the Kennicutt or a similar IMF for disk star formation.  The use
of a Kennicutt or a similar IMF for disk star formation is supported
by many observations and models of solar neighbourhood stars,
including a semi-analytic model by \citet{no03} in which a
Salpeter-like IMF provides good agreement with the observed
metallicities of solar-neighbourhood stars.  We have also confirmed
that while using the Arimoto-Yoshii IMF \citep{ay87} with $x=1$ for
both star formation modes substantially increases metallicities
compared with the Kennicutt IMF, a combination of the Kennicutt IMF
for quiescent star formation and the Arimoto-Yoshii IMF for starbursts
does not reach the observed metallicities.

\section{ICM ABUNDANCES IN HIGH-REDSHIFT CLUSTERS}

\subsection{Evolution of [Fe/H]}

\begin{figure}
\epsfxsize=\hsize
\epsfbox{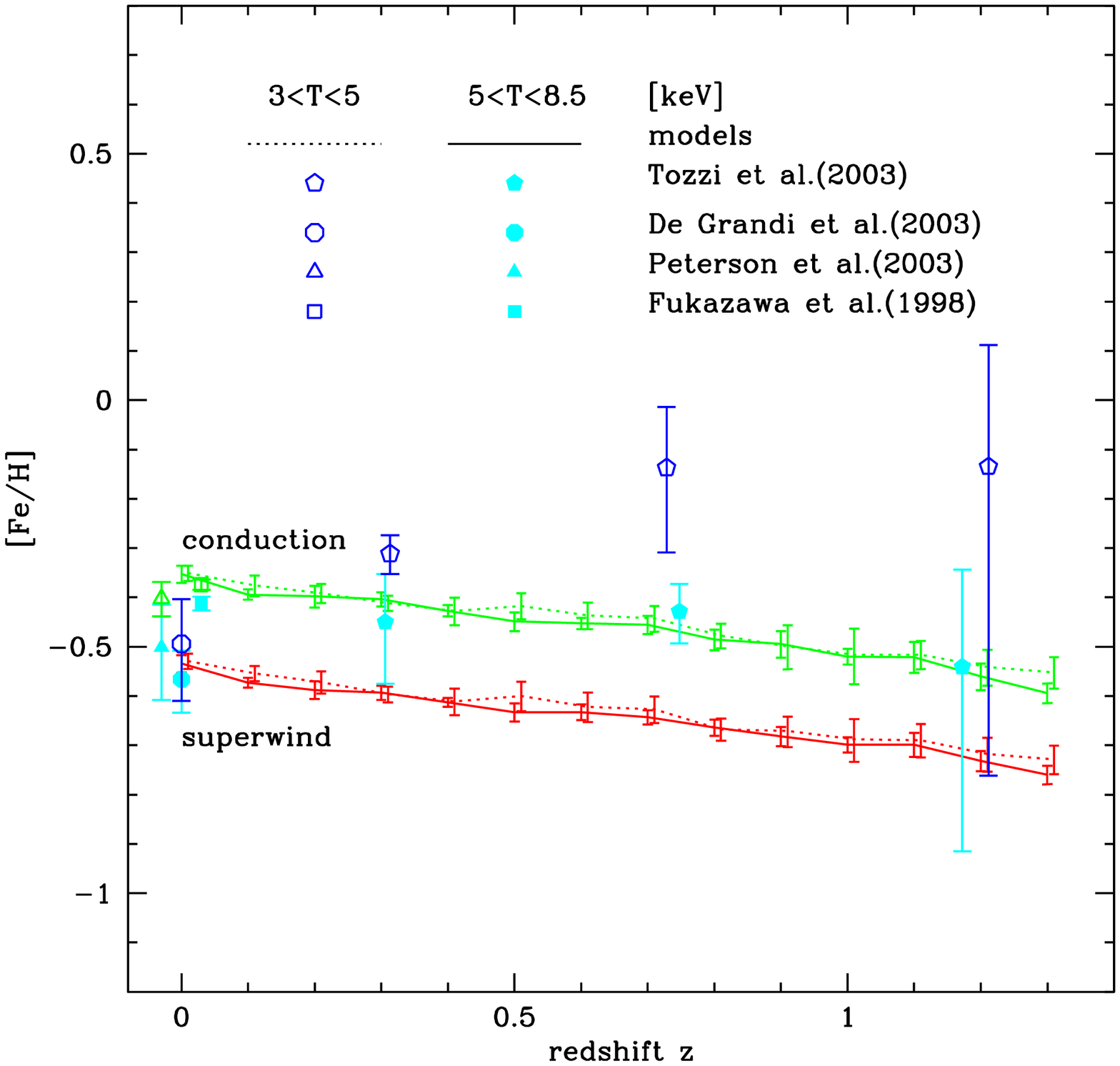}

\caption{The redshift evolution of [Fe/H] in the ICM. The lines show
the mean abundances predicted by the models for two different
temperature ranges, 3-5keV (dotted) and 5-8.5keV (solid), for both the
superwind and conduction models (lower and upper pairs of lines, as
labelled). The error bars on the lines show the 1$\sigma$ scatter.
The symbols show the observal data, with means and 1$\sigma$ scatter
computed in each bin. Pentagons indicate the data for high redshift
clusters from \citet{t03}, circles, triangles and squares are mean
metallicities of clusters at $z=0$ given by \citet{dG03}, \citet{p03}
and \citet{f98}, respectively.  The symbols for $z=0$ are given small
offsets for clarity. The observational data are binned into the same
temperature ranges as the models.}

\label{fig:Fez}
\end{figure}

With the advent of the new generation of $X$-ray satellites,
measurements of ICM metallicities are now becoming possible in
high-redshift clusters, thus allowing direct tests of ICM
evolution. In Fig.~\ref{fig:Fez}, we show what our models predict for
the evolution of the iron abundance in the ICM with redshift in the
range $0<z<1.3$. At each redshift, we calculate the mean Fe abundance
for clusters in two different ranges of temperature, $3<T<5$keV and
$5<T<8.5$keV. We show results for the standard conduction and
superwind models. In the same figure we also show observational data
from \citet{t03}, based on measurements of the iron abundances in 18
distant clusters with $0.3\la z\la 1.3$ using {\it Chandra} and {\it
XMM-Newton}. We have binned this data in redshift and in the same
ranges of temperature as used for the models. Note that we have no
information about metallicity gradients for these distant clusters, so
we have not applied any corrections to the measured abundances to get
global values. We also show observational data for $z=0$ clusters
binned in the same ranges of temperature. The $z=0$ data are corrected
for metallicity gradients as previously described.

Our models predict a weak decline in the Fe abundance with increasing
redshift, by about 30\% over the range $0<z<1.3$, and furthermore this
decline is predicted to be the same for both low and high temperature
clusters. \citet{t03} concluded from their own data that the evolution
in Fe abundance with redshift is negligible, but we see that the data
for the hotter ($5<T<8.5$keV) clusters are also quite consistent with
the weak decline predicted by our models. The data for the cooler
($3<T<5$keV) clusters suggest higher Fe abundances and a weak increase
with redshift relative to the hotter clusters, both of which would
conflict with our model, but the uncertainties are very large.

\subsection{Evolution of [O/Fe]}

\begin{figure}
\epsfxsize=\hsize
\epsfbox{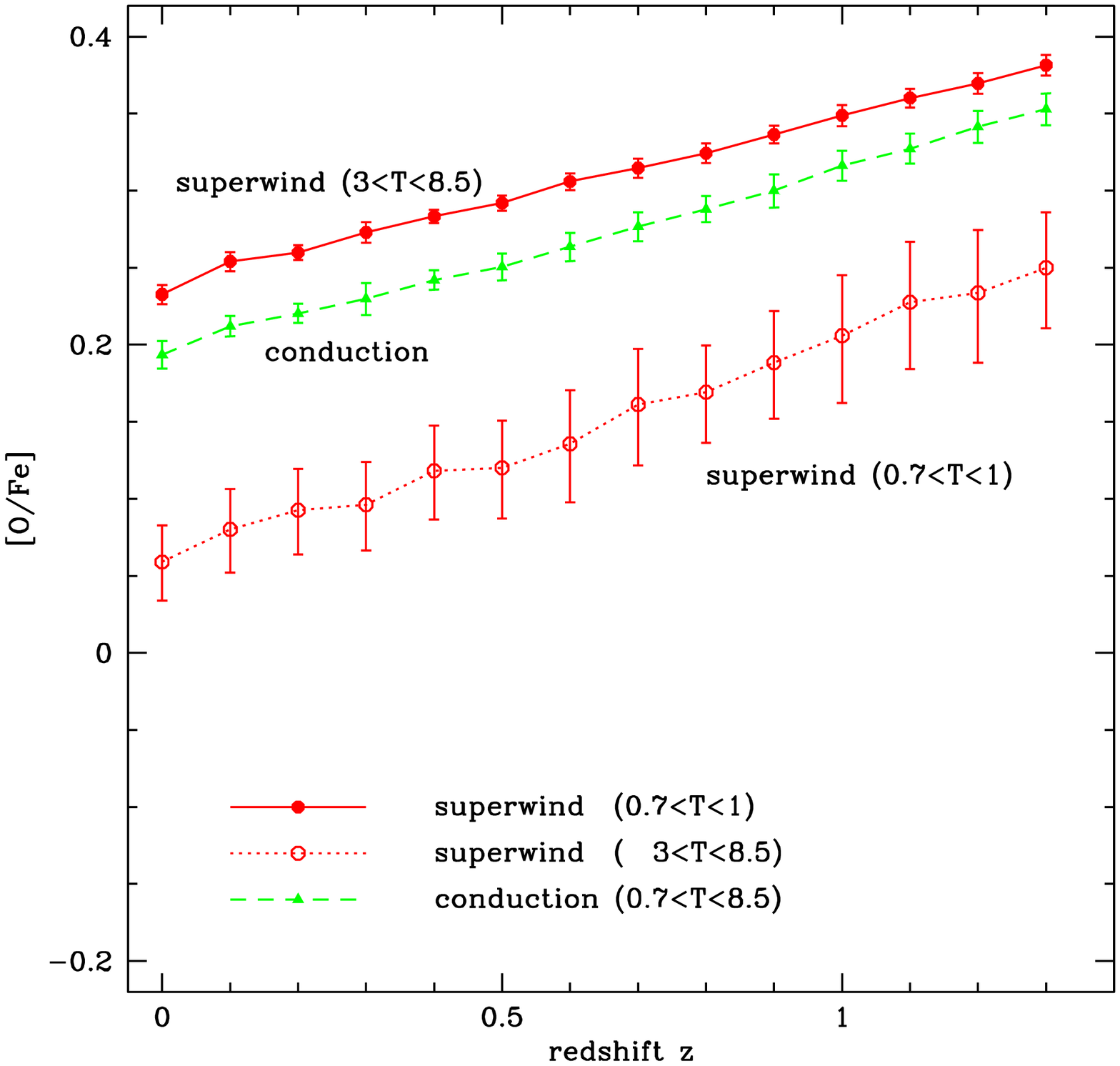}

\caption{The redshift evolution of [O/Fe] in the ICM.  The lines
 show the predictions for the conduction and superwind models
 respectively.  In the case of the superwind model, the sample is
 divided into two temperature ranges: high (3-8.5 keV; {\it solid
 line}) and low (0.7-1 keV; {\it dotted line}) temperatures.  }

\label{fig:OFez}
\end{figure}

The abundances of elements other than Fe have not yet been measured in
high-redshift clusters, so we simply present predictions for the
evolution of O abundances from our model. For clusters selected in the
same temperature range at different redshifts, the mean O abundance is
predicted to decline with redshift even more weakly than for Fe, by
only $\sim 10\%$ in over the range $0<z<1.3$. The difference in
behaviour is caused by the difference in the timescales for enrichment
by SNe~Ia and SNe~II. The O/Fe ratio in the ICM is therefore predicted
to increase with redshift. This is shown in Fig.~\ref{fig:OFez}. For
the superwind model, there is a significant difference in O/Fe between
low (0.7-1 keV) and high (3-8.5 keV) temperature clusters, caused by
recapture of superwind ejecta in clusters with $T\gtrsim 1.1$keV.

\begin{figure*}
\epsfxsize=\hsize
\epsfbox{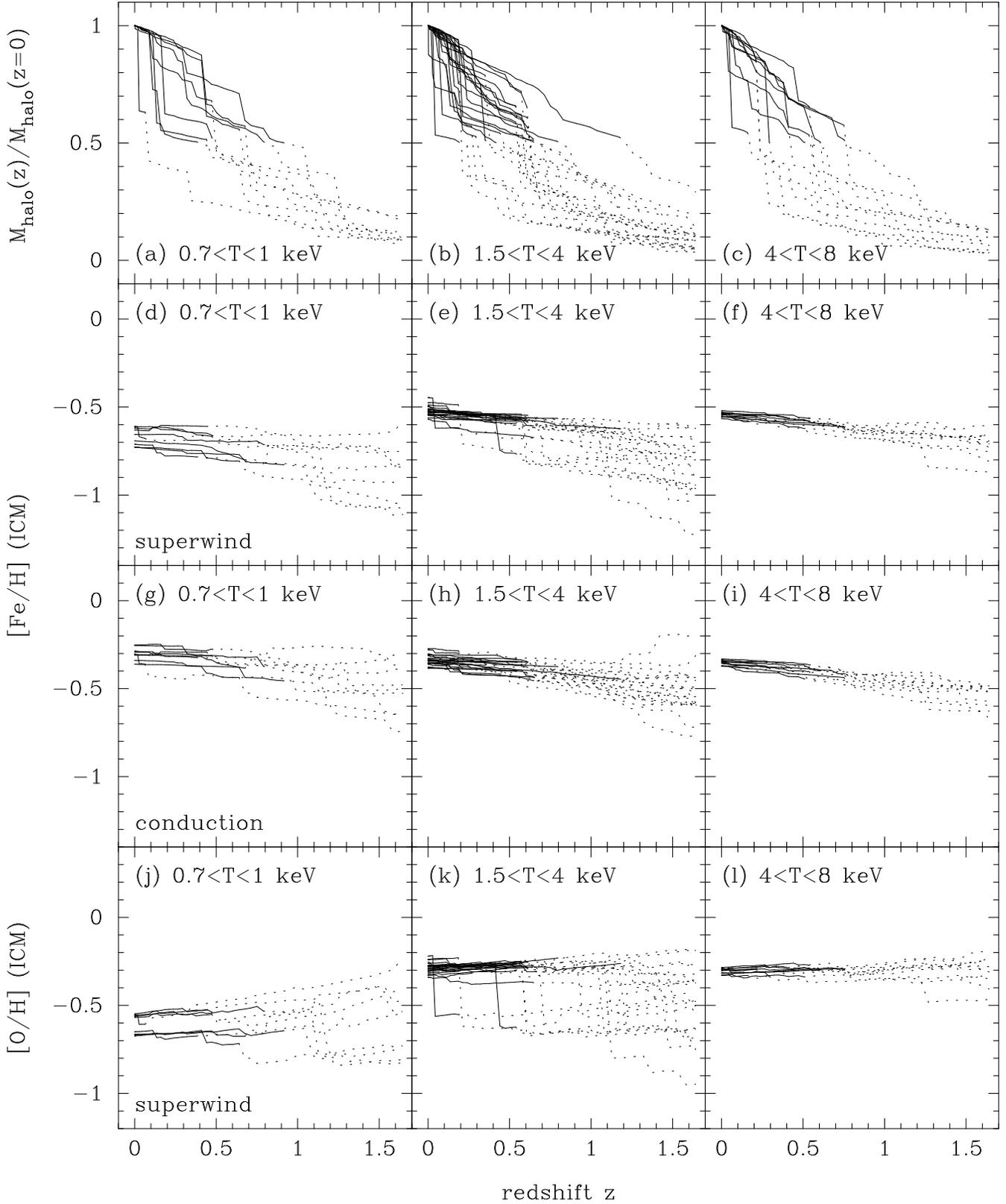}

\caption{The evolution of the main progenitors of present-day
clusters, as defined in the text. The left, middle and right panels in
each row show the evolution of clusters with low (0.7-1 keV),
intermediate (1.5-4 keV) and high (4-8 keV) present-day temperatures
respectively. Top row: The evolution of halo mass of the main
progenitor, relative to the present-day cluster mass.  Second and
third rows: The evolution of iron abundance [Fe/H] in the ICM of the
main progenitor, in the superwind and conduction models
respectively. Fourth row: The evolution of the oxygen abundance [O/H]
in the ICM of the main progenitor, in the superwind model.  In all
panels, the line style changes from dotted to solid once the formation
redshift of the cluster is passed.  }

\label{fig:hist1}
\end{figure*}

\section{METAL ENRICHMENT HISTORIES}

In this section, we investigate what our model predicts for the
enrichment histories of individual clusters. For this purpose, we
divide clusters into three populations based on their {\em
present-day} ICM temperatures: low (0.7-1 keV), intermediate (1.5-4
keV) and high (4-8 keV), corresponding to cluster masses $3.6-6.2
\times 10^{13}$, $1.1-5.0 \times 10^{14}$ and $0.5-1.4 \times 10^{15}
h^{-1}M_{\odot}$ respectively.  These ranges are chosen to illustrate
different behaviours regarding recapture of gas by halos in the
superwind model. For the low temperature clusters, a significant
quantity of metals and gas have been expelled from the progenitors of
these halos without recapture.  The intermediate temperature clusters
experience recapture of gas and metals at low redshifts, so we can see
how the recapture process affects the evolution of the ICM
metallicity.  The high temperature clusters have already recaptured
gas and metals at high redshifts.

In Fig.~\ref{fig:hist1} we show the evolution of the halo mass and ICM
metallicity for the main progenitors of present-day clusters, for a
randomly chosen set of clusters in each range of present-day
temperature. We define the main progenitor of a present-day cluster at
any earlier epoch as follows \citep{lc93}: starting from the
present-day halo, we follow the halo merger history back in time, and
at each merger event, we follow the branch in the merger tree
corresponding to the more massive progenitor involved in that merger
event. We define the formation redshift $z_{f}$ of a cluster as the
earliest redshift at which the main progenitor has at least half of
the present cluster halo mass \citep{lc93}. For $z\leq z_{f}$, the
{\it main} progenitor is also the {\it most massive} progenitor, but
at earlier times this may not be the case.

The top row of panels in Fig.~\ref{fig:hist1} shows the evolution of
the halo mass of the main progenitor relative to the present-day halo
mass, illustrating the spread in cluster assembly histories. We plot
each cluster trajectory in Fig.~\ref{fig:hist1} as a solid line for
$z<z_f$ and as a dotted line for $z>z_f$. In the $\Lambda$CDM model,
clusters typically form around $z\sim 0.5$, but with a wide
dispersion, and with more massive clusters typically forming more
recently. The medians and the 10\% and 90\% percentiles of the
distribution of $z_{f}$ values are (0.64,0.38,0.95) for low
temperature, (0.57,0.27,0.79) for medium temperature, and
(0.46,0.21,0.75) for high temperature clusters respectively.

The lower panels of Fig.~\ref{fig:hist1} show the evolution of the ICM
abundances in the main progenitor of each cluster, for the same set of
cluster halo assembly histories as in the top row. The second and
third rows show the Fe abundance in the superwind and conduction
models, while the fourth row shows the O abundance in the superwind
model only. The Fe abundances typically evolve only weakly with
redshift over the range $0<z<1.5$, which is similar to the behaviour
seen in Fig.~\ref{fig:Fez}.  The dispersion in the final metallicity
and in the metallicity evolution is larger in the lower temperature
clusters. For the intermediate temperature clusters, the dispersion in
the metallicity evolution is larger in the superwind model than in the
conduction model, due to the ejection and then recapture of gas in the
former case. These recapture events can be seen as abrupt upward jumps
in metallicity in panel~(e). The evolution of the O abundance in the
superwind model shows a somewhat larger dispersion than the evolution
in Fe abundance. The jumps in O metallicity due to recapture of
superwind material are larger than for Fe (compare panels (k) and
(e)), because the superwinds are preferentially enriched in products
of SNe~II rather than SNe~Ia. Some of the evolution tracks for the O
abundance show a {\em decline} in mean ICM abundance with {\em
increasing} time. This behaviour can be understood as follows: the
dispersion in O abundance between halos of the same mass increases
with decreasing halo mass, due to variations in the number of $\sim
L_{*}$ galaxies within a halo resulting from the stochastic nature of
the halo assembly and galaxy formation processes. As the halo mass
builds up, the metallicity tends to converge towards the average value
for clusters with that present-day temperature: progenitor halos with
above-average ICM metallicity tend to merge with other halos having
lower metallicity, and the resulting mixing of the hot gas components
produces the declining tracks seen for instance in panel~(j).

\begin{figure*}
\epsfxsize=\hsize
\epsfbox{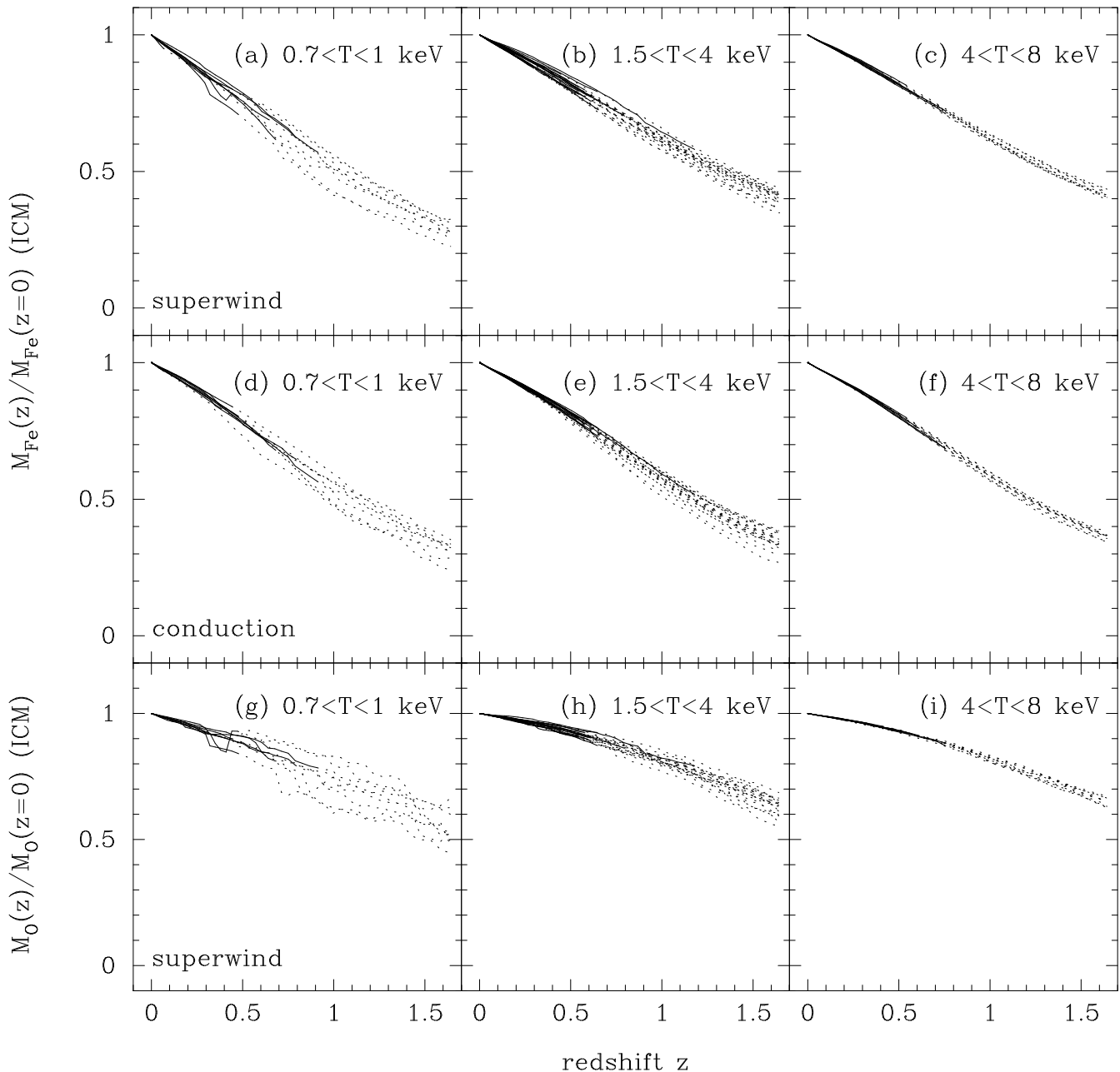}

\caption{(a)-(f): The evolution of the total mass of metals in the
  ICM, summed over all progenitors, relative to the present-day metal
  mass. Metals ejected by superwinds are included if these are
  recaptured by the present-day. The top two rows show the evolution
  of the iron mass in the superwind and conduction models
  respectively, while the bottom row shows the evolution of the oxygen
  mass in the superwind model only. The different columns show the
  results for different ranges of present-day cluster temperature in
units of keV as indicated by the labels.}

\label{fig:hist2}
\end{figure*}

In Fig.~\ref{fig:hist2} we show the build-up of the total mass of
metals found in the present-day ICM for the same set of clusters as in
Fig.~\ref{fig:hist1}. We calculate these evolutionary tracks by
summing over the ICM metal masses of all of the progenitors of a
present-day cluster. We include in this sum metals which have been
ejected by superwinds, if these metals have been recaptured by the
present day. These tracks thus allow us to determine the redshifts at
which metals in the ICM of the present-day cluster were actually
produced (in the sense of being ejected from stars in supernova
explosions). The top two rows of panels show the evolution of the iron
mass in the superwind and conduction models, for different ranges of
present-day cluster temperature. The bottom row shows the evolution of
the oxygen mass in the superwind model. Most of the metals are
typically produced well before the clusters are assembled, with the
oxygen being produced even earlier than the iron. For example, in the
superwind model, half of the O is produced by redshifts
$z_O=(1.8,2.1,2.1)$ for the low, medium and high temperature clusters
respectively, compared to $z_{Fe}=(1.0,1.3,1.3)$ for producing half of
the Fe. The corresponding redshifts in the conduction model are not
much different: $z_O=(1.7,1.8,1.9)$ and $z_{Fe}=(1.1,1.2,1.2)$. Thus,
the ICM enrichment occured mostly in many smaller progenitors of a
present-day cluster. We also see that metal production typically
occured slightly earlier in high mass than low mass clusters, even
though high mass clusters were assembled (in the sense of reaching
half of their present mass) slightly later.

We can compare our results for the redshift of metal production in
clusters with the results from the semi-analytical model of
\citet{del04}. De~Lucia et al. use the instantaneous recycling
approximation throughout, so their results can be considered to apply
to oxygen, but are not valid for iron, for which the delayed
production by SNe~Ia is important. Furthermore, they only compute the
ICM evolution for the single example of a $10^{15} h^{-1}M_{\odot}$
cluster. For three different feedback/ejection schemes, they calculate
that 60-80\%, 35-60\%, and 20-45\% of the metals are incorporated into
the ICM at redshifts larger than 1, 2, and 3, respectively. Note that
in their definition, metals are ``incorporated'' into the ICM either
when they are ejected from galaxies directly into the ICM, or when
metals ejected from the cluster are recaptured. Comparing with the
predictions for high temperature clusters in our own models, we find
that the fraction of oxygen incorporated into the ICM at redshifts
larger than 1, 2 and 3 is 47\%, 23\%, and 10\% respectively for the
superwind model, and 79\%, 51\%, and 24\% for the conduction model.
Thus in our superwind model, O is incorporated into the ICM later than
in the models of De~Lucia et al., while in our conduction model, the
incorporation happens at similar redshifts. (Note that the ``metal
incorporation'' redshifts given here are different from the ``metal
production'' redshifts derived from Fig.~\ref{fig:hist2}, since the
latter include ejected metals among the metals produced by a given
redshift, provided these ejected metals have been recaptured by the
present day.) Since we assume the same $\Lambda$CDM cosmology as
De~Lucia et al., the differences must arise from using somewhat
different assumptions about gas cooling, star formation, feedback and
gas ejection.

Mergers of halos are expected to lead to extensive mixing of the ICM,
which will tend to erase any metallicity gradients. The prediction of
the models that typically most of the iron in the ICM is produced well
before half of the cluster mass has been assembled then seems to make
it harder to understand the iron abundance gradients seen in ``cooling
flow'' clusters. However, since the production of iron by SNe~Ia is
long-delayed, it may still be possible to build up iron abundance
gradients in the central regions of clusters by ejection of iron from
the central galaxy after most of the cluster mass has been
assembled. This problem needs to be investigated using gas-dynamical
simulations of cluster formation.

We have also calculated the build-up of metals within cluster
galaxies, counting metals in both stars and in the cold gas. When we
compute trajectories for the evolution of the total mass in metals in
galaxies summed over all cluster progenitors, relative to the present
metal mass, we find that there is very little dispersion between
different clusters of the same mass (much less than for the ICM metal
mass shown in Fig.~\ref{fig:hist2}). The trajectories are also very
similar for different ranges of cluster temperature, and for the superwind
and conduction models. For clusters in the temperature range
$0.7<T<8$keV, half of the O mass in cluster galaxies is produced by
redshifts $z_O=2.6-2.8$ for the superwind model, and by $z_O=2.7-2.9$
in the conduction model. For the Fe mass in cluster galaxies, the
corresponding redshifts are $z_{Fe}=1.7-1.8$ in the superwind model,
and $z_{Fe}=1.9-2.0$ in the conduction model. The metals in cluster
galaxies are thus produced significantly earlier than the metals in
the ICM.

\begin{figure}
\epsfxsize=\hsize
\epsfbox{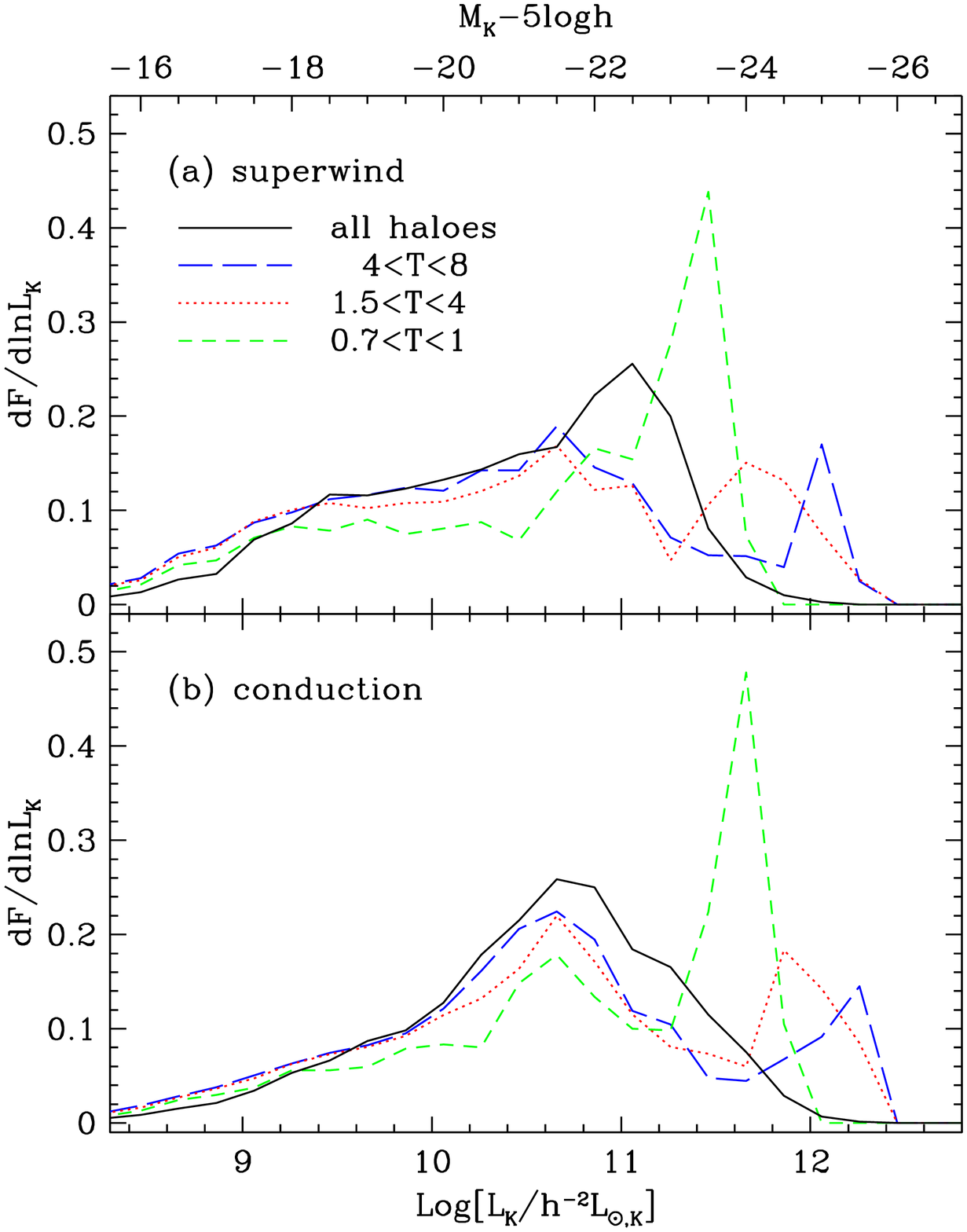}

\caption{The fractional contribution of galaxies to iron enrichment as
 a function of $K$-band luminosity, $L_{K}$. Iron in all baryonic
 components (stars, cold gas, hot gas or ejected gas) is included, and
 the distributions are normalized to unit area. Panel (a) shows
 results for the superwind model and (b) for the conduction model.
 The long-dashed, dotted, and short-dashed lines indicate high,
 intermediate, and low temperature clusters, respectively.  The solid
 lines show the distributions averaged over halos of all masses.}

\label{fig:metaldist}
\end{figure}

Finally, we examine which galaxies contribute the most to cluster
metal enrichment. Fig.~\ref{fig:metaldist} shows the total amount of
iron which galaxies have produced (either retained within the galaxy,
injected into the ICM, or ejected from the cluster) as a function of
their $K$-band luminosities at $z=0$. The contribution of each galaxy
is calculated by summing the iron production over its formation
history. The distributions over luminosity ${\rm d}F(\leq L_{K})/{\rm
d}\ln L_{K}$ are then normalized to unit area.  Results are shown for
both the superwind and conduction models, and also split according to
present-day cluster temperature. The solid line in each panel shows
the iron production averaged over halos of all masses (including halos
much smaller than clusters). These distributions are quite similar to
the distributions of total stellar mass, which is expected since the
mass of iron produced is closely related to the mass of stars
formed. We also find that the normalized distributions for the mass of
iron injected into the ICM are very similar to the distributions in
Fig.~\ref{fig:metaldist} for the total mass of iron produced.

The distributions of iron production in Fig.~\ref{fig:metaldist} for
the clusters show pronounced peaks at the high luminosity end. As the
cluster mass or temperature increases, the position of the peak moves
up in luminosity, but its amplitude decreases. This peak is produced
by the central galaxies in the cluster halos, and reflects the
distribution of total stellar mass \citep{bfbcl03}. The peak is most
prominent for the lowest temperature clusters, because mergers are
most effective at concentrating the stellar mass into a single central
galaxy in such systems. Since the position of the peak due to central
galaxies depends on halo mass, it is smeared out when we average over
all halos, producing a smoother distribution (see the solid lines in
Fig.~\ref{fig:metaldist}).

We find that in the superwind model, half of the iron is produced by
galaxies with luminosities $L_K$ brighter than $1.1\times 10^{11}$,
$4.2\times 10^{10}$ and $3.6\times10^{10} h^{-2}L_{\odot}$ for low,
medium and high temperature clusters respectively, corresponding to
galaxy stellar masses $5.4\times 10^{10}$, $2.2\times 10^{10}$ and
$1.9\times10^{10} h^{-1} M_{\odot}$ respectively. The median galaxy
masses and luminosities for oxygen production are almost identical to
those for iron.  In the conduction model, the median galaxy masses and
luminosities for metal production are somewhat higher: for iron
production, the median K-band luminosities are $1.2\times 10^{11}$,
$6.2\times 10^{10}$ and $5.4\times 10^{10} h^{-2}L_{\odot}$, and the
median stellar masses are $7.3\times 10^{10}$, $3.5\times 10^{10}$ and
$3.1\times 10^{10} h^{-1} M_{\odot}$ respectively, with very similar
values for oxygen production. When we look at metal production
averaged over halos of all masses, we find median luminosities $L_K$
for producing Fe or O of $4.2\times 10^{10}$ and $5.5\times 10^{10}
h^{-2}L_{\odot}$ respectively for the superwind and conduction models,
and median stellar masses of $1.8\times 10^{10}$ and $2.5\times 10^{10}
h^{-1} M_{\odot}$ respectively.  We thus see that the median K-band
luminosity (or stellar mass) for metal production is similar (within a
given model) for high mass clusters and field galaxies, but
significantly larger in low mass clusters.  For comparison, the
characteristic luminosity of field galaxies in the K-band is
$L_{K*}=6\times 10^{10} h^{-2}L_{\odot}$ \citep{c01}. The metal
production in clusters and in the field is thus predicted to be
dominated by galaxies with $L_K \gtrsim L_{K*}$.

In our models, the median galaxy luminosities and masses for metal
injection into the ICM are almost identical to those for metal
production.  Our median stellar masses for metal production are 2--3
times larger in the case of the most massive clusters than the median
galaxy mass of $1\times 10^{10} h^{-1} M_{\odot}$ for metal
(effectively oxygen) injection into the ICM calculated by
\citet{del04} for a $10^{15} h^{-1}M_{\odot}$ cluster. However, we are
in agreement De Lucia et al. on the general conclusion that the ICM
metal enrichment is dominated by fairly massive galaxies, in contrast
to suggestions \citep[e.g.][]{garnett02} that metal ejection by dwarf
galaxies has dominated the chemical enrichment of the ICM.

\section{SUMMARY}

We have investigated the metal enrichment of the ICM for clusters of
different masses using a semi-analytical model of galaxy formation and
evolution, {\sc galform} (Cole et al. 2000).  We have implemented in
this model a detailed treatment of metal production by both SNe~Ia and
SNe~II, based on the predictions of stellar evolution models for both
stellar lifetimes and yields of different heavy elements. This enables
us to follow the evolution of different elements (O, Mg, Si and Fe),
including the long time delays for SNe~Ia. Our model consistently
calculates the cycling of gas and metals between stars, cold gas and
hot gas. In order to reproduce the observed present-day galaxy
luminosity function using a realistic baryon fraction in the models,
we have considered two alternative mechanisms to prevent the
overproduction of bright galaxies: (a) galactic superwinds which expel
gas and metals from dark halos, with subsequent recapture in very
massive halos, and (b) thermal conduction which transfers heat from
the outer to the central regions of hot gas halos, thereby supressing
gas cooling.  We considered two different IMFs for bursts of star
formation: the Kennicutt IMF that we also adopt for quiescent disk
star formation, and a top-heavy, flat IMF with $x=0$.  We have found
that the form of the IMF is strongly constrained by the metal
abundances of O, Mg, Si and Fe in the ICM. Our main conclusions can be
summarized as follows:

\begin{enumerate}
\renewcommand{\theenumi}{(\arabic{enumi})}
 \item A top-heavy IMF in starbursts is required in order to obtain
       ICM metal abundances as high as those observed, when the IMF
       for quiescent star formation in disks is fixed to be a
       standard, bottom-heavy IMF, such as the Kennicutt IMF.  A burst
       IMF with $x=0$ reproduces the observed metal abundances in the
       ICM.  We have also tried the Arimoto-Yoshii IMF with $x=1$ in
       bursts, and confirmed that it fails to produce sufficient
       metals.  Considering that the IMF for quiescent star formation
       has been well constrained by many studies, including a
       semi-analytical model \citep{no03}, this is strong evidence in
       support of a top-heavy IMF with $x\simeq 0$ in starbursts.

\item Both the superwind and conduction models reproduce the observed
       ICM metal abundances and abundance ratios.  The superwind model
       predicts lower metal abundances in low temperature ($T\la 1$
       keV) clusters. Therefore if the drop in [Fe/H] for clusters
       with $T\la 1$ keV found using {\it ASCA} is real, this could be
       evidence for the action of superwinds.  The current data are
       still limited, however, so future observations, particularly of
       low temperature clusters, would strongly constrain the
       superwind model.

\item Our model is in broad agreement with the observed ratio of ICM
       gas mass to the total $b_{\rm J}$-band stellar luminosity in
       clusters, $M_{\rm hot}/L_{b_{\rm J}}$.  In particular, the
       superwind model predicts a decrease of $M_{\rm hot}/L_{b_{\rm
       J}}$ in low temperature clusters, as is observed, because the
       superwind expels gas from these clusters.

\item The iron metallicity, [Fe/H], is predicted to decrease only
       slowly with increasing redshift.  The difference in [Fe/H]
       between $z=0$ and 1 is at most 0.2 dex.  This mild evolution is
       broadly consistent with the recent observations by \citet{t03}.
       In contrast with [Fe/H], [O/H] is almost independent of
       redshift over this range, so [O/Fe] increases with redshift.

\item The metal production histories of clusters having the same
      present-day mass are very similar. In contrast, the cluster mass
      assembly histories show a wide variety. This suggests that the
      variety of metallicity gradients observed in clusters may be
      caused not only by observational errors but also by the variety
      of assembly histories. Presumably this is the origin of the
      diversity in the observed metallicities of present-day clusters,
      because most observations measure metallicities only at cluster
      centres.

\item In clusters and in the field, metal production is dominated by
      galaxies with K-band luminosities similar to or brighter than
      $L_{*}$. In low-mass clusters, the contribution of the central
      galaxy to metal production is very important, but this effect is
      weaker in high-mass clusters.

\end{enumerate}

Throughout this paper, we have neglected any contribution of
intracluster stars to the metal enrichment of the ICM, since in our
model all stars are in galaxies.  Based on recent measurements of the
intracluster light, it has been suggested that intracluster stars
might have produced a large fraction of the iron observed in the ICM
\citep{zgz04}. The intracluster stars would presumably originate from
material tidally stripped from galaxies or ejected during violent
mergers, so the luminosity function of galaxies in clusters would also
be affected. In our model, tidal stripping of stars from galaxies
would not affect the total mass of iron produced (if all other
parameters remain the same), but it might affect the fraction of the
iron which is ejected into the ICM. This possibility clearly deserves
further observational and theoretical study.

It is also evident that, in order to better constrain our model, we need
more accurate ICM metallicity measurements for larger samples of
clusters. These measurements need to cover the outer parts of clusters
as well as the central regions, in order derive accurate global
metallicities.  The observational situation will be substantially
improved by the {\it Astro-E2} satellite scheduled for launch in 2005,
as well as by further data from {\it XMM-Newton} and {\it
Chandra}. Further constraints on the model can be obtained by examining
the abundances of different heavy elements in galaxies, and we will
investigate this latter topic in future papers.

\section*{ACKNOWLEDGMENTS}    
We thank Takashi Okamoto, Richard Bower, Stefano Ettori, Simon
D. M. White, Yasushi Fukazawa and Takeshi Go Tsuru for useful
discussions.  We also acknowledge support from the PPARC rolling grant
for extragalactic astronomy and cosmology at Durham.  MN is supported by
the Japan Society for the Promotion of Science for Young Scientists
(No.207).  CMB is supported by a Royal Society University Research
Fellowship.

\appendix
\section{Correction of the observational data for metallicity gradients}
Some clusters are known to have substantial metallicity gradients.  The
so-called `cooling flow' clusters show enhancements in the metallicities
of some elements toward the cluster centre \citep[e.g.,][]{tbkfm01,
mfb03, dG03}.  \citet{tbkfm01, tkhbp04} have reported that in cooling
flow clusters O and Mg among the $\alpha$ elements do not show the
metallicity gradients, whereas Fe and Si tend to show gradients, at
least from the cluster centre to around half of the virial radius,
despite the fact that Si is an $\alpha$ element.  Non-cooling flow
clusters generally do not show metallicity gradients for any elements.
Our model predicts the global metallicity of the ICM, averaged over the
whole cluster. Therefore we need to correct the observed metallicities
to obtain values averaged over the whole cluster, because most
observations give only central values.

\citet{dG03} compiled Fe metallicity gradients in 12 cooling flow
clusters.  Based on their results, we assume that all cooling flow
clusters have the same shape of metallicity gradients in both Fe and
Si,
\begin{equation}
 Z(r)=Z_{0}\left[1+\left(\frac{x}{x_{c}}\right)^{2}\right]^{-\alpha},
\label{eqn:metalgrad}
\end{equation}
where $x=r/r_{200}$, $r_{200}$ is the radius within which the mean
overdensity is 200 times the critical density, $x_{c}=0.04$ and
$\alpha=-0.18$.  We define the average metallicity within radius $r$
as,
\begin{equation}
 \bar{Z}(\leq r) = \frac{\int_{0}^{r}Z(r)\rho_{g}(r)r^{2}dr}
{\int_{0}^{r}\rho_{g}(r)r^{2}dr},
\end{equation}
where $\rho_{g}(r)$ is the gas density.  We assume that the observed
metallicities correspond to average values within some radius $r$,
$\bar{Z}(\leq r)$, and then use the assumed form of the metallicity
profile (equation ({eqn:metalgrad}))to correct these to mean values
$\bar{Z}(\leq r_{\rm vir})$ within the virial radius $r_{\rm vir}$.
The central metallicity $Z_{0}$ is thus treated as a parameter chosen
so as to match an observed metallicity $\bar{Z}(\leq r)$.

\subsection{De Grandi et al.'s sample}
\citet{dG03} give the iron masses of individual clusters as a
function of $\Delta$, which is a spherical overdensity,
\begin{equation}
 \Delta = \frac{3M_{\rm tot}(<r_{\Delta})}{4\pi \rho_{c,z}r_{\Delta}^{3}},
\end{equation}
where $M_{\rm tot}(<r)$ is the total mass within a radius $r$ and
$\rho_{c,z}$ is the critical density of the universe at redshift $z$.
Using the data on the distributions of total and gas mass given by
\citet{edm02}, we obtain an averaged metallicity for each cluster
within $r_{\Delta}$, $\bar{Z}(\leq r_{\Delta})$.  The gas density
distribution, $\rho_{g}(r)$ is assumed to follow an isothermal profile
with a core,
\begin{equation}
 \rho_{g}(r)\propto\left[1+\left(\frac{r}{r_{s}}\right)^{2}\right]^{-1}.
\end{equation}
To obtain $r_{\Delta}$, we use the structural parameters given by
\citet{edm02} assuming an NFW profile \citep{nfw97} for the total
density distribution.  We have found that the use of structural
parameters assuming a King profile hardly affects averaged
metallicities.  $r_{s}$ is assumed to be the same as the scale radius
of the NFW profile, but the results are not sensitive to the exact
value chosen.

The metallicity at the cluster centre, $Z_{0}$ in
eq.(\ref{eqn:metalgrad}), depends on the individual cluster.  Using
the above equations and the values at $r_{2500}$, at which the
observed data are given in the above papers, we obtain the average
metallicity $\bar{Z}$ within the virial radius ($r\leq r_{\rm vir}$,
corresponding to $\Delta\simeq 100$ in a $\Lambda$CDM universe),
\begin{equation}
 \bar{Z}=\frac{M_{{\rm Fe},2500}}{M_{{\rm gas},2500}}\frac{\bar{Z}(\leq
 r_{\rm vir})}{\bar{Z}(\leq r_{2500})}.
\end{equation}
The average [Fe/H] is easily estimated from $\bar{Z}$.

\subsection{Peterson et al.'s sample}
Peterson et al. (2003) observed only the central parts of clusters
within a $1'$ angular radius.  Their observed metallicities of Fe and
Si are assumed to correspond to $\bar{Z}(\leq r_{1'})$, where $r_{1'}$
is the radius at which the angular size is equal to 1 arcmin. The
average metallicity within the virial radius is then obtained from
\begin{equation}
 \bar{Z}=Z_{\rm obs}\frac{\bar{Z}(\leq r_{\rm vir})}{\bar{Z}(\leq r_{1'})}.
\end{equation}
The parameters regarding redshift and structure required to obtain averaged
metallicities are tabulated in their paper.

Recent observations suggest that the metallicity gradients of Si are
very similar to those of Fe \citep{tbkfm01, tkhbp04, mfb03}.
Therefore we assume that Si has the same shape of metallicity profile
as Fe.  As already mentioned, because O and Mg seem not to have any
metallicity gradients, we regard the observed values of central
metallicities as equivalent to the metallicities averaged over the
whole cluster.

\subsection{Fukazawa et al.'s sample}
The measurements of Fe and Si abundances by Fukazawa et al. (1998)
include data from the outer parts of the clusters. We therefore assume
that their abundance measurements are close to global averages, and do
not apply any corrections.

\subsection{Baumgartner et al.'s sample}
\citet{blhm03} obtained average abundances of several elements,
including Fe and Si, in bins of cluster temperature by performing a
stacking analysis of observations of 273 clusters from the {\it ASCA}
archives. Although their cluster sample is large, metallicity gradients
combined with the large point spread function for {\it ASCA} cause
problems for interpreting their results, especially for Fe. The iron
abundance is usually estimated from the Fe-K or Fe-L lines. For
high-temperature clusters, the photon flux in the Fe-K line is larger
than that in the Fe-L line, and the Fe-K line yields accurate
measurements of the iron abundance. On the other hand, for low and
intermediate temperature clusters, the Fe-K line becomes very weak, and
the photon flux is larger in the Fe-L line, so the Fe-L line dominates
in abundance estimates. Lower temperature gas emits more photons in the
Fe-L line than in the Fe-K line. This has the consequence that for
clusters with cooling cores, the observationally-estimated iron
abundance is weighted towards the value in the low temperature
component, that is, the central region, where the iron abundance is
typically higher than the average over the whole cluster. Thus, as
\citet{f98} have done, the data for the central region should be removed
to get a fairer estimate of the metallicity averaged over the whole
cluster.  \citet{blhm03} have not done this, and this may be the reason
why their data show an enhancement of the iron metallicity at
intermediate cluster temperatures.  For this reason, we do not include
their data on Fe abundances in our comparison with the models. Their
data on Si abundances seem to be less affected by this problem, so we do
include those data in our comparison.

\bsp

\begin{thebibliography}{}   
\bibitem[Anders \& Grevesse(1989)]{ag89}Anders E., Grevesse N., 1989,
				  Geochim. Cosmochim. Acta, 53, 197
\bibitem[Arimoto \& Yoshii(1987)]{ay87}Arimoto N., Yoshii Y., 1987,
                                 A\&A, 173, 23
\bibitem[Baugh et al. (2004a)]{baugh04a} Baugh C.M., et al. (the 2dFGRS team), 
2004a, MNRAS, 351, L44
\bibitem[Baugh et al. (2004b)]{baugh04b} Baugh C.M., Lacey C.G., Frenk C.S. 
Granato G.L. Silva L., Bressan A., Benson A.J., Cole S., 
2004b, MNRAS, in press (astro-ph/0406069) (B04)
\bibitem[Baumgartner et al.(2003)]{blhm03}Baumgartner W.H., Loewenstein
				  M., Horner D.J., Mushotzky R.F., 2003,
				  ApJ, submitted (astro-ph/0309166)
\bibitem[Bell et al.(2003)]{bmkw03}Bell E.F., McIntosh D.H., Katz N.,
				  Weinberg M.D., 2003, ApJ, 149, 289
\bibitem[Benson et al.(2003a)]{bbflbc03}Benson A. J., Bower R.G., Frenk C. S., 
			      Lacey C. G., Baugh C. M., Cole S., 
			      2003a, ApJ, 599, 38
\bibitem[Benson et al.(2003b)]{bfbcl03}
Benson A.J., Frenk C.S., Baugh C.M., Cole S., Lacey C.G.,
2003b, MNRAS 343, 679
\bibitem[Cole(1991)]{c91}Cole S., 1991, ApJ, 367, 45
\bibitem[Cole et al.(1994)]{cafnz94}Cole S., Aragon-Salamanca A., Frenk
                                  C. S., Navarro J. F., Zepf
                                  S. E., 1994, MNRAS, 271, 781
\bibitem[Cole et al.(2000)]{clbf00}Cole S., Lacey C. G., Baugh C. M.,
                                  Frenk C. S., 2000, MNRAS, 319, 168 (CLBF)
\bibitem[Cole et al.(2001)]{c01}Cole S. et al. 2001, MNRAS, 2001, 326, 225
\bibitem[David, Forman \& Jones(1991)]{dfj91}David L.P., Forman W.,
				  Jones C., 1991, ApJ, 380, 39
\bibitem[De Grandi \& Molendi(2002)]{dG02}De Grandi S., Molendi S.,
				  2002, ApJ, 567, 163
\bibitem[De Grandi et al.(2004))]{dG03}De Grandi S., Ettori S.,
				  Longhetti M., Molendi S., 2004, A\&A,
				  419, 7
\bibitem[De Lucia, Kauffmann \& White (2004)]{del04}De Lucia, G., 
                                   Kauffmann, G. \& White, S.D.M., 2004, MNRAS, 349, 1101
\bibitem[Eke et al.(1996)]{eke96}
Eke V.R., Cole S., Frenk C.S., 1996, MNRAS, 282, 263
\bibitem[Enoki, Nagashima \& Gouda(2003)]{eng03}Enoki M., Nagashima M.,
				  Gouda N., 2003, PASJ, 55, 133
\bibitem[Ettori, De Grandi \& Molendi(2002)]{edm02}Ettori S., De Grandi
				  S., Molendi S., 2002, A\&A, 391, 841
\bibitem[Fukazawa et al.(1998)]{f98}Fukazawa Y., Makishima K., Tamura
				  T., Ezawa H., Xu H., Ikebe Y., Kikuchi
				  K., Ohashi T., 1998, PASJ, 50, 187
\bibitem[Garnett (2002)]{garnett02}Garnett, D.R., 2002, ApJ, 581, 1019
\bibitem[Gibson \& Matteucci(1997a)]{gm97a}Gibson B.K., Matteucci F.,
				  1997a, MNRAS, 291, L8
\bibitem[Gibson \& Matteucci(1997b)]{gm97b}Gibson B.K., Matteucci F.,
				  1997b, ApJ, 475, 47
\bibitem[Granato et al.(2004)]{granato04}
Granato G.L., De Zotti G., Silva L., Bressan A.,
Danese L., 2004, ApJ, 600, 580.
\bibitem[Greggio \& Renzini(1983)]{gr83}Greggio L., Renzini A., 1983,
				  A\&A, 118, 217
\bibitem[Grevesse \& Sauval(1998)]{gs98}Grevesse N., Sauval A.J., 1998, Space
				 Sci. Rev., 85, 161
\bibitem[Hatton et al. (2003)]{hatton03}
Hatton, S., Devriendt, J.E.G., Ninin, S., Bouchet, F.R., Guiderdoni, B., 
Vibert, D., 2003, MNRAS, 343, 75
\bibitem[Helly et al.(2003)]{h03}
Helly J.C., Cole S., Frenk C.S., Baugh C.M., Benson A.J., 
Lacey C.G., 2003, MNRAS, 338, 903
\bibitem[Huang et al.(2003)]{hgct03}Huang J.-S., Glazebrook K., Cowie
				  L.L., Tinney C., 2003, ApJ, 584, 203
\bibitem[Irwin \& Bregman(2001)]{ib01}Irwin J.A., Bregman J.N., 2001,
				  ApJ, 546, 150
\bibitem[Jenkins et al.(2001)]{j01}Jenkins A., Frenk C. S., White
			      S. D. M., Colberg J. M., Cole S.,
			      Evrard A. E., Couchman H. M. P.,
			      Yoshida N. 2001, MNRAS, 321, 372
\bibitem[Kauffmann (1996)]{k96}Kauffmann G., 1996, MNRAS, 281, 475
\bibitem[Kauffmann \& Charlot(1998)]{kc98}Kauffmann G., Charlot S.,
				 1998, MNRAS, 294, 705
\bibitem[Kauffmann \& Haehnelt(2000)]{kh00}Kauffmann G., Haehnelt M.,
				  2000, MNRAS, 311, 576
\bibitem[Kauffmann, White \& Guiderdoni(1993)]{kwg93}Kauffmann G.,
                                  White S. D. M., Guiderdoni
                                  B., 1993, MNRAS, 264, 201
\bibitem[Kennicutt(1983)]{k83}Kennicutt R.C., 1983, ApJ, 272, 54
\bibitem[Kroupa(2002)]{k02}Kroupa P., 2002, Sci., 295, 82
\bibitem[Lacey \& Cole (1993)]{lc93}
Lacey C.G., Cole S., 1993, MNRAS 262, 627
\bibitem[Lacey et al.(2005)]{l05}Lacey C.G., et al. 2005, in preparation
\bibitem[Larson(1969)]{l69}Larson R.B., 1969, MNRAS, 145, 297
\bibitem[Lia, Portinari \& Carraro(2002)]{lpc02}Lia C., Portinari L.,
				 Carraro G., 2002, MNRAS, 330, 821
\bibitem[Loewenstein \& Mushotzky(1996)]{lm96}Loewenstein M., Mushotzky
				  R.F., 1996, ApJ, 466, 695
\bibitem[Maoz \& Gal-Yam(2004)]{mg04}Maoz D., Gal-Yam A., 2004, MNRAS,
				  347, 951
\bibitem[Marigo(2001)]{m01}Marigo P., 2001, A\&A, 370, 194
\bibitem[Matsushita, Ohashi \& Makishima(2000)]{mom00}Matsushita K.,
				  Ohashi T., Makishima K., 2000, PASJ,
				  52, 685
\bibitem[Matsushita, Finoguenov \& B\"{o}hringer(2003)]{mfb03}Matsushita
				  K., Finoguenov A., B\"{o}hringer H.,
				  2003, A\&A, 401, 443
\bibitem[Matteucci \& Gibson(1995)]{mg95}Matteucci F., Gibson B.K.,
				  1995, A\&A, 304, 11
\bibitem[Matteucci \& Greggio(1986)]{mg86}Matteucci F., Greggio L.,
				 1986, MNRAS, 154, 279
\bibitem[Matteucci \& Fran\c{c}ois(1989)]{mf89}Matteucci F.,
				 Fran\c{c}ois P., 1989, MNRAS, 239, 885
\bibitem[Menci et al.(2002)]{mcfgp02}Menci N., Cavaliere A., Fontana A.,
				  Giallongo E., Poli F., 2002, ApJ, 575, 18
\bibitem[Moretti, Portinari \& Chiosi(2003)]{mpc03}Moretti A., Portinari
				  L., Chiosi C., 2003, A\&A, 408, 431
\bibitem[Mushotzky et al.(1996)]{m96}Mushotzky R., Loewenstein M.,
				  Arnaud K.A., Tamura T., Fukazawa Y.,
				  Matsushita K., Kikuchi K., Hatsukade
				  I., 1996, ApJ, 466, 686
\bibitem[Nagashima \& Gouda(2001)]{ng01}Nagashima M., Gouda N., 2001,
				 MNRAS, 325, L13
\bibitem[Nagashima \& Okamoto(2004)]{no03}Nagashima M., Okamoto T.,
				  2004, submitted (astro-ph/0404486)
\bibitem[Nagashima et al.(2001)]{ntgy01}Nagashima M., Totani T., Gouda
                                 N., Yoshii Y., 2001, ApJ, 557, 505
\bibitem[Nagashima et al.(2002)]{nytg02}Nagashima M., Yoshii Y., Totani T.,
				 Gouda N., 2002, ApJ, 578, 675
\bibitem[Nagashima \& Yoshii(2004)]{ny03}Nagashima M., Yoshii Y.,
				 2004, ApJ, 610, 23
\bibitem[Nagashima et al.(2005)]{n05}Nagashima M., et al. 2005, in preparation
\bibitem[Navarro, Frenk \& White(1997)]{nfw97}Navarro J.F., Frenk C.S.,
				  White S.D.M., 1997, ApJ, 490, 493
\bibitem[Nomoto et al.(1997)]{n97}Nomoto K., Iwamoto K., Nakasato N.,
				  Thielemann F.-K., Brachwitz F.,
				  Tsujimoto T., Kubo Y., Kishimoto N.,
				  1997, Nuclear Physics, A621, 467c
\bibitem[Norberg et al. (2002)]{norberg02}
Norberg P., et al. (the 2dFGRS team), 2002, MNRAS, 336, 907
\bibitem[Okoshi et al.(2004)]{ongy04}Okoshi K., Nagashima M., Gouda N.,
				 Yoshioka S., 2004, ApJ, 603, 12
\bibitem[Pagel \& Tautvai\v{s}ien\.{e}(1995)]{pt95}Pagel B.E.J.,
				 Tautvai\v{s}ien\.{e}
				 G., 1995, MNRAS, 276, 505
\bibitem[Peacock et al. (2001)]{peacock01}
Peacock J.A., et al. (the 2dFGRS team), 2001, Nature, 410, 169.

\bibitem[Percival et al. (2002)]{percival02}
Percival W.J., et al. (the 2dFGRS team), 2002, MNRAS, 337, 1068

\bibitem[Peterson et al.(2001)]{peterson01}
Peterson J.R., et al., 2001, A\&A, 365, L104
\bibitem[Peterson et al.(2003)]{p03}Peterson J.R., Kahn S.M., Paerels
				  F.B.S., Kaastra J.S., Tamura T.,
				  Bleeker J.A.M., Ferrigno C., Jernigan
				  J.G., 2003, ApJ, 590, 207
\bibitem[Portinari, Chiosi \& Bressan(1998)]{pcb98}Portinari L., Chiosi
				  C., Bressan A., 1998, A\&A, 334, 505
\bibitem[Press \& Schechter(1974)]{ps74}Press W., Schechter P., 1974, ApJ, 187, 425
\bibitem[Renzini et al.(1993)]{renzini93}
Renzini A., Ciotti L., D'Ercole A., Pellegrini S., 1993, 
ApJ, 419, 52.
\bibitem[Salpeter(1955)]{s55}Salpeter E.E., 1955, ApJ, 121, 161
\bibitem[Sanderson et al.(2003)]{spflm03}Sanderson A.J.R., Ponman T.J.,
				  Finoguenov A., Lloyd-Davies E.J.,
				  Markevitch M., 2003, MNRAS, 340, 989
\bibitem[Sanderson \& Ponman(2003)]{sp03}Sanderson A.J.R., Ponman T.J.,
				  2003, MNRAS, 345, 1241
\bibitem[Smith \& Gallagher(2001)]{sg01}Smith L.J., Gallagher III J.S.,
				  2001, MNRAS, 326, 1027
\bibitem[Somerville \& Primack(1999)]{sp99}Somerville R.S.,
                                  Primack J. R., 1999, MNRAS, 310, 1087
\bibitem[Somerville, Primack \& Faber(2001)]{spf01}Somerville R.S.,
                                  Primack J. R., Faber S.M., 2001,
                                  MNRAS, 320, 504
\bibitem[Spergel et al. (2003)]{wmap03}
Spergel D.N., et al. (the WMAP Team),  2003, ApJS, 148, 175
\bibitem[Takahara \& Takahara(1981)]{tt81}Takahara M., Takahara F.,
				  1981, PTP, 65, 369
\bibitem[Tamura et al.(2001)]{tbkfm01}Tamura T., Bleeker J.A.M., Kaastra
				  J.S., Ferrigno C., Molendi S., 2001,
				  A\&A, 379, 107
\bibitem[Tamura et al.(2004)]{tkhbp04}Tamura T., Kaastra J.S., den
				  Herder J.W.A., Bleeker J.A.M.,
				  Peterson J.R., 2004, A\&A, 420, 135
\bibitem[Thomas(1999)]{t99}Thomas D., 1999, MNRAS, 306, 655
\bibitem[Thomas, Greggio \& Bender(1998)]{tgb98}Thomas D., Greggio L.,
				  Bender R., 1998, MNRAS, 296, 119
\bibitem[Thomas \& Kauffmann(1999)]{tk99}Thomas D., Kauffmann G., 1999,
				 in Spectrophotometric dating of stars
				 and galaxies, ed. I. Hubeny, S. Heap \&
				 R. Cornett, Vol. 192 (ASP Conf. Ser.), 261
\bibitem[Timmes, Woosley \& Weaver(1995)]{tww95}Timmes F.X., Woosley
				  S.E., Weaver T.A., 1995, ApJS, 98, 617
\bibitem[Tornatore et al.(2004)]{tbmrt04}Tornatore L., Borgani S.,
				  Matteucci F., Recchi S., Tozzi P.,
				  2004, MNRAS, 349, 19
\bibitem[Tozzi et al.(2003)]{t03}Tozzi P., Rosati P., Ettori S., Borgani
				  S., Mainieri V., Norman C., 2003, ApJ,
				  593, 705
\bibitem[Tsujimoto et al.(1995)]{tnyhyt95}Tsujimoto T., Nomoto K.,
				 Yoshii Y., Hashimoto M., Yanagida Y.,
				 Thielemann F.-K., 1995, MNRAS, 277, 945
\bibitem[Tsuru et al.(1997)]{takp97}Tsuru T.G., Awaki H., Koyama K.,
				  Ptak A., 1997, PASJ, 49, 619
\bibitem[Valdarnini(2003)]{v03}Valvarnini R., 2003, MNRAS, 339, 1117
\bibitem[Whelan \& Iben (1973)]{wi73} Whelan, J., Iben, I., 1973, ApJ,
  186, 1007 
\bibitem[White \& Frenk (1991)]{wf91} White, S.D.M. \& Frenk, C.S.,
1991, ApJ, 379, 52
\bibitem[Yoshida et al.(2002)]{y02}
Yoshida N., Stoehr F., Springel V., White S.D.M., 
2002, MNRAS, 335, 762
\bibitem[Yoshii, Tsujimoto \& Nomoto(1996)]{ytn96}Yoshii Y., Tsujimoto
				 T., Nomoto K., 1996, MNRAS, 462, 266
\bibitem[Zaritsky, Gonzalez \& Zabludoff(2004)]{zgz04}Zaritsky D.,
				  Gonzalez A.H., Zabludoff A.I., 2004,
				  ApJ, in press (astro-ph/0406291)
\bibitem[Zepf \& Silk(1996)]{zs96}Zepf S.E., Silk J., 1996, ApJ, 466, 114
\end{thebibliography}
\end{document}